\documentclass[aps,showpacs,prd,superscriptaddress,a4paper,nofootinbib]{revtex4}

\usepackage{latexsym}
\usepackage{amsmath}
\usepackage{amssymb}
\usepackage{ulem}
\usepackage{fancyhdr}
\usepackage{natbib}
\usepackage{url}
\usepackage{graphicx}
\usepackage{grffile}
\usepackage[vcentering,dvips]{geometry}
\usepackage{sidecap}

\begin{document}

\title{Observational cosmology using characteristic numerical
relativity }

\author{P.~J. van der Walt}
\affiliation{
Department of Mathematics, Rhodes University, Grahamstown 6140, South
Africa
}
\author{N.~T. Bishop}
\affiliation{
Department of Mathematics, Rhodes University, Grahamstown 6140, South
Africa
}

\begin{abstract}

The characteristic formalism in numerical relativity, which has been
developed to study gravitational waves, and the observer metric
approach in observational cosmology both make use of coordinate
systems based on null cones. In this paper, these coordinate systems
are compared and it is then demonstrated how characteristic
numerical relativity can be used to investigate problems in
observational cosmology. In a numerical experiment using the
characteristic formalism, it is shown how the historical evolution
of a LTB universe compares to that of the $\Lambda$CDM model
given identical observational data on a local observer's past null
cone. It is demonstrated that, at an earlier epoch of the LTB model, the
observational data would not be consistent with that of the $\Lambda$CDM
model.

\end{abstract}
\maketitle

\section{Introduction}

Provided that General Relativity is valid up to the largest scales,
under ideal circumstances, the space-time structure of the Universe
can be determined by solving the full set of Einstein field
equations (EFEs) using observational data as the boundary
conditions. Such an observational approach makes use of minimal
assumption and observations as the boundary conditions dictate the
outcome of the solution. Because of difficulties in obtaining and
interpreting observational data, the conventional approach to
cosmology is rather based on parameterized models where a model is
assumed and then validated against different sets of observations
and constraints. Assuming a homogeneous universe with a non-zero
cosmological constant and with suitable adjustments of parameters,
predictions using the $\Lambda$CDM model show remarkable correlation
with observed cosmological parameters. The assumption of
homogeneity, which is an integral part of this approach, has the
implication that local results from simple models can easily be
extrapolated to the whole Universe.

However, doing investigations such as quantifying homogeneity on
different scales, testing the verifiability of cosmology
\cite{ellis85}, validating the Copernican principle \cite{uzan08}
and determining the metric of the Universe \cite{hellaby09}, a more
suitable methodology is one where the EFEs are solved using a
general metric and boundary conditions derived from observations.
Since homogeneity is not assumed, conclusions following from the
observational approach are necessarily limited to the causally
connected region in the interior of our current past null cone on
which observations are located. Since the EFEs are solved in a
general form, or at least more general than the $\Lambda$CDM model,
there are more degrees of freedom and also more required boundary
conditions. Sufficiently accurate and complete observations do not
currently exist, however, it is anticipated that the next generation
of astronomical surveys will begin to provide enough detail to
perform observational solutions in spherical symmetry.

Much of the development on observational cosmology was inspired by
the seminal paper of Kristian \& Sachs \cite{kristian66} published
in 1966. In their work, observational data was used to determine
a general metric but since they made use of series expansions, it
was necessarily restricted to a region close to the observer. The
ideas introduced by Kristian \& Sachs were the starting point of
further developments by Ellis, Stoeger, Maartens and others in their
\emph{Observational Cosmology} programme with its initial
publication in 1985 \cite{ellis85}. In these developments,
observational coordinates, based on the concept introduced by Temple
in 1938 \cite{temple38}, were implemented to extend the region
investigated by Kristian \& Sachs to higher redshifts. Observational
coordinates are based on null geodesics on the past null cone (PNC)
and are therefore the natural framework on which electromagnetic
radiation reaches an observer. Solving the EFEs for cosmology then
consists of two problems: firstly, astronomical observations are
used to determine the metric on the local PNC and secondly these
form the final values of a characteristic final value (CFVP) problem
which determines the historical evolution of the region causally
connected to the PNC (i.e. the interior of the PNC). The causally
connected region is of fundamental importance since it defines the
limits on which cosmological models can be validated from direct
observations.

In developments towards exact solutions in observational cosmology,
spherical symmetrical solutions are taken as a first step to refine
the methods (see for instance \cite{stoeger92, araujo00, araujo09}).
Spherical symmetry by itself is not necessarily unrealistic since
the Universe does appear to be highly isotropic. The spherical
symmetrical inhomogeneous Lema\^{i}tre-Tolman-Bondi (LTB) model in
observational coordinates is therefore an important tool for
verifying these developments. The LTB model in its standard form
also provides a useful framework to investigate solutions from
direct observations. In this approach, observations on the PNC are
transformed to cosmological coordinates and then related to the
coefficients of the LTB metric (e.g. see \cite{musta98}). These
transformations require numerical solutions to handle observational
data and recent work by Lu \& Hellaby \cite{lu07} and McClure \&
Hellaby \cite{mcclure09} developed and refined methods for setting
up the local PNC from realistic observations with the intention to be
implemented on data obtainable in the next generation of
astronomical surveys. It should be noted that the emphasis in
observational approaches lies in the fact that the initial (final)
conditions are provided by direct observations on the PNC and that
the EFEs are implemented in a general form with minimal assumptions.

In numerical relativity (NR) methods to implement general solutions
of the EFEs have been well established for strong gravitational
scenarios where gravitational waves are expected to be formed.
Similar to electromagnetic radiation, gravitational radiation also
propagates along null geodesics and null cones also provide a
natural frame of reference. In this case, however, it is the future
null cone that is of interest in the form of a characteristic
initial value problem (CIVP). The characteristic formalism in NR is
based on the theoretical developments of Bondi, van der Burg \&
Metzer \cite{bondi62} and Sachs \cite{sachs62} which were part of
the \emph{Gravitational Waves in General Relativity} programme
initiated by Bondi in the late 1950s. These developments were of
fundamental importance to the understanding of gravitational waves
and with the advancement of computational technology, it was
recognised that the characteristic formalism holds several
computational benefits. Among these are: the fact that the EFEs
simplify to ordinary differential equations along characteristics,
which are less expensive to compute and the conformal method,
developed by Penrose \cite{penrose63}, can be used to represent
infinity on a finite grid, making modelling asymptotical behaviour
possible in full non-linearity. One drawback of a null
cone coordinate system is its behaviour around caustics where it can
become multi-valued and singular. In astrophysical problems, this problem is
usually avoided by combining the characteristic formalism with the
Cauchy formalism where the latter is used to solve regions where
caustics are expected while the former is used to extract the
solution in the far field \cite{bish93,Reisswig:2009us}. Treating caustics directly
on the null cone has also been investigated \cite{friedrich83} but
this approach has not been implemented numerically. A comprehensive
overview of characteristic numerical relativity can be found in
\cite{winicour09}.

In the context of cosmology, the CIVP in NR poses the advantages
that cosmological developments can leverage on previous developments
in astrophysical problems, the causally connected region of the PNC
can be computed using realistic boundary conditions free of
constraints and the coordinates are simpler than observational
coordinates. The simplicity of the coordinates does however
introduce restrictions in that the location where the PNC
refocusses, the apparent horizon (AH), becomes multi-valued. The
importance of the AH has been emphasised in \cite{hellaby06} and
\cite{araujo09} and methods to cross this region have been presented
in \cite{lu07}, \cite{mcclure09} and \cite{yoo08} and similar
approaches will probably work for the characteristic formalism as
well. Although recognising the importance of the AH, the initial
steps taken in the current work will be limited to the region where
the characteristic formalism is well behaved while the region where
caustics occur will be postponed for future research. Applying the
CIVP in NR to observational cosmology has previously been
investigated by Bishop \& Haines in \cite{bish96} and the work
presented in this paper is a continuation of their work. Recent work
by Hellaby \& Alfedeel \cite{hellaby09} defined an algorithm to
implement an approach based on observational coordinates with the
intention of a numerical implementation. In their work,
considerations were taken into account for passing the AH and some
other details not taken into account in this paper but a numerical
implementation of their work has not been presented yet.

The work presented in this paper is about evolution off the
null cone i.e. the second part of the observational cosmology
problem. We develop a computer code, the input to which is data on
the PNC obtainable, in principle, directly from observations. The code
uses the EFEs to evolve the model into the past. The code reported here is
a first step towards the construction of a general code for observational
cosmology, and is limited to the case of spherical symmetry.

The code is used to investigate an issue in cosmology. The standard model
of the universe is $\Lambda$CDM, but it is well known that the observational
data also fits a LTB model. The code is used with data on the PNC
generated by the $\Lambda$CDM model, and we evolve into the past using the
$\Lambda=0$ EFEs. The result is that we compute the data that would be observed
on the PNC at an earlier epoch within the context of the LTB model. We then
ask the question: Could this data also be interpreted as being that of some
$\Lambda$CDM model? The answer is ``No'', which indicates that, if the universe
is LTB, then not only are we at a special position in space, but also that the
the present time is special in the history of the universe.

In section \ref{s-Charac}, the CIVP is formulated and related to
cosmology. Section \ref{s-Num} presents the implementation of the
code, and section \ref{s-Codev} describes its verification. The
results of running the code with $\Lambda$CDM model data are given
in section \ref{s-LTB}. The paper ends with a Conclusion, section
\ref{s-Conc}.

\section{Characteristic formalism}
\label{s-Charac}
\subsection{Metric and coordinates}

The essence of the characteristic formalism is a frame of reference
based on outgoing null cones that evolve from values on an initial
null cone. The idea is conceptualised in figure \ref{sec:obs-coords}.
$G$ is a timelike geodesic, and $u$ is the proper time on $G$.
Null geodesics emanating from $G$ have constant $(u,\theta,\varphi)$,
and near $G$ the angular coordinates $\theta$ and
$\varphi$ have the same meaning as in spherical polar coordinates.
The coordinate $r$ is the diameter, or area, distance from the cone
vertex, which means that the surface area of a shell of constant $r$ is
$4\pi r^2$.

A spherically symmetric null cone metric based on the Bondi-Sachs
metric is: \footnote{The notation used here is based on that of
\cite{bish97} and substituting $W=V-r$ will give the original
notation of Bondi in \cite{bondi60}.}:
\begin{align}
ds^2 = -e^{2\beta} \left(1+\frac{W}{r} \right) du^2
        - 2 e^{2\beta} dudr + r^2 \{ d \theta ^2
        +  \sin ^2 \theta d \varphi ^2\} . \label{sec:civp-met}
\end{align}
The hypersurface variables, $\beta$ and $W$, represent the deviation
from a Minkowskian null cone and the coordinate system is defined
such that $\beta$ and $W$ vanish at the vertex of each null cone
i.e.: at $r=0$, equation (\ref{sec:civp-met}) reduces to a
Minkowskian metric.

Substituting (\ref{sec:civp-met}) into the EFEs, using the form
$R_{ab} = 8\pi(T_{ab} - \frac{1}{2} T g_{ab})$, with the
stress-tensor for a dust-like fluid ($T_{ab} = \rho v_a v_b$ and
$T=\rho$), leads to expressions for $\beta$ and $W$:
\begin{align}
\beta_{,r} &= 2\pi r \rho (v_{1})^{2} \label{sec:civp-beta}\\
W_{,r} &= e^{2 \beta} - 1 - 4 \pi e^{2 \beta} \rho r^2
\label{sec:civp-W}
\end{align}
with the initial conditions inherent to the coordinate definition as
$\beta(0) = W(0) = 0$. Further, substituting the dust stress tensor
and (\ref{sec:civp-met}) into the conservation equation,
$T^{ab}_{\;\;;b}=0$, it follows that:
\begin{align}
v_{1,u}  & = \frac{1}{v_{1}} \left\{
    (2 v_1 v_0 - V_w (v1)^{2}) \beta_{,r}
    + (V_{w} v_{1} - v_0) v_{1,r}
    + \frac{1}{2}(v_1) ^{2} V_{w,r} \right\} \label{sec:civp-v1} \\
\rho_{,u}  & = \frac {1}{v_1 } \left\{ \rho \left[ V_w \left(
\frac{2v_1}{r} + v_{1,r} \right)
    - \left(  \frac{2v_{0}}{r} + v_{0,r}  \right)
    + V_{w,r} v_1 \right]
    + \rho_{,r} \left( V_w v_1 - v_0 \right)
    - \rho v_{1,u} \right\} \label{sec:civp-rho} \\
\textrm{with:~~} V_w & = 1+\frac{W}{r} . \nonumber
\end{align}
An equation for $v_{0,u}$ is also obtained from this but making use
of the normalisation condition, $g^{ab}v_{a}v_{b}=-1$, a direct
expression for $v_{0}$ in terms of $v_{1}$, $\beta$ and $W$ can be
written as:
\begin{align}
v_{0} &= \frac{1}{2} v_1 V_w + \frac{1}{2} e^{2 \beta} v_1^{-1}.
\label{sec:civp-v0}
\end{align}

Without any cosmological considerations, having the values on the
initial null cone for $\rho$ and $v_1$, equations
(\ref{sec:civp-beta}) to (\ref{sec:civp-v0}) forms a hierarchical
system that can be solved in the order (\ref{sec:civp-beta}),
(\ref{sec:civp-W}) and (\ref{sec:civp-v0}), then solving equations
(\ref{sec:civp-v1}) and (\ref{sec:civp-rho}) evolves the system to
the next null cone where the process can be repeated until the
domain of calculation has been covered. Since these equations are
all dependent on each other an iterative scheme is required for a
numerical solution.

\subsection{Cosmological coordinates}

\begin{figure}[h!]
\begin{center}$
\begin{array}{ll}
\hspace{0pt}
\includegraphics[width=0.35\textwidth]{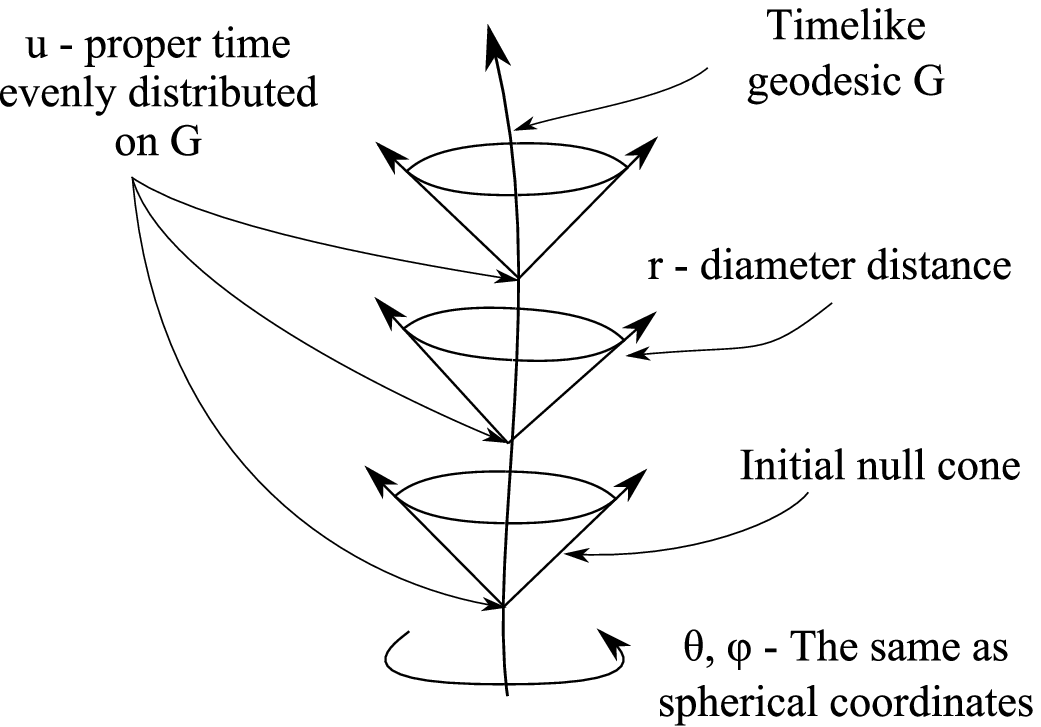} &
\hspace{10pt}\includegraphics[width=0.35\textwidth]{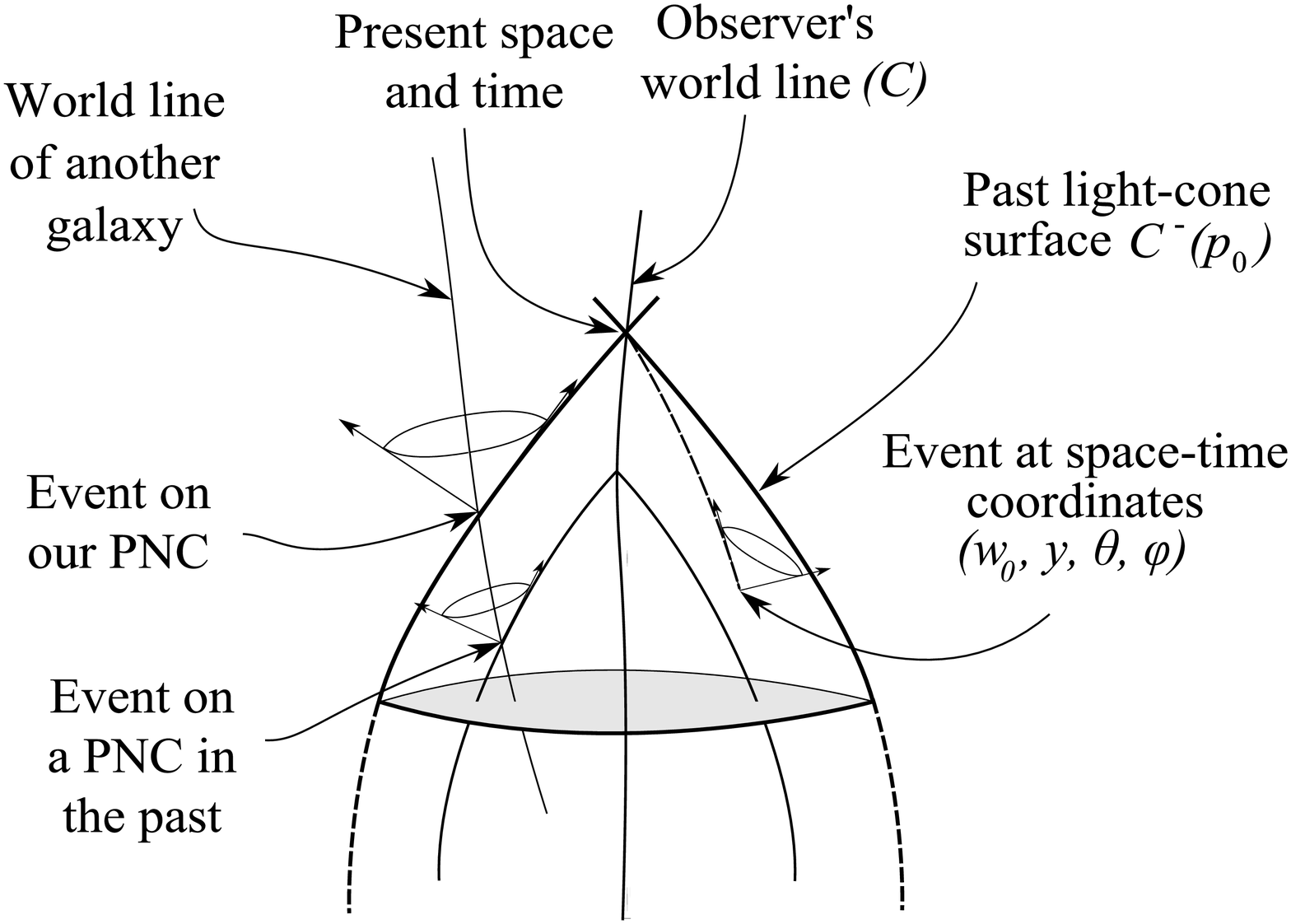}
\end{array}$
\end{center} \vspace{0pt} \caption{On the left hand side, the null
coordinates of the characteristic formalism and on the right, the
null cone coordinates of observational coordinates. }
\label{sec:obs-coords}
\end{figure}

In observational cosmology, the observer metric is similar to
(\ref{sec:civp-met}) but allows for a more general treatment of the
radial coordinate. Using the notation of \cite{stoeger92}, the
metric is defined as:
\begin{align}
ds^2 = -A^2(w,y) dw^2
        + 2 A(w,y) B(w,y) dwdy + C^2(w,y) \{d \theta ^2
        + \sin ^2 \theta d \varphi ^2\} . \label{sec:obs-met}
\end{align}
With reference to the right hand part of figure
\ref{sec:obs-coords}, the coordinates $x^i=\{w,y,\theta, \varphi\}$
are defined as (see \cite{ellis85} and \cite{stoeger92}): $w$, the
proper time coordinate on an observer's world line ($C$), such that
$w=constant$ hypersurfaces are past light cones. The instance $w_0$
is the current time of a local observer, defining our PNC. $y$, is a
radial coordinate defined as a null geodesic on the past light cone
comoving with $w$. There is some freedom in the in the
interpretation of $y$ which includes: an affine parameter, the
redshift $z$ and the diameter distance $C(w,y)$. As with the
characteristic formalism, $\theta$ and $\varphi$ are the spherical
inclination and azimuth angles respectively

The objective is then to determine $A(w_0,y)$, $B(w_0,y)$ and
$C(w_0,y)$ from directly observable quantities and then to solve a
CFVP to determine the interior of the null cone. For the spherically
symmetrical dust case observable quantities can be limited to: the
redshift $z$, the apparent luminosity $l$ and number counts $n$. The
redshift ($z$) is measured in terms of distance related to $l$ and
$n$ is obtained in terms of $z$. These quantities are then related
through source evolution functions to the absolute luminosity $L$
and mass per source $\mu$. An important aspect here is that source
evolution functions should be model independent \cite{musta98}.

There are three important differences between (\ref{sec:civp-met})
and (\ref{sec:obs-met}). Firstly, the characteristic formalism is
more restricted in the radial coordinate $r$ which is chosen to be
the diameter distance ($r=y=C(w,y)$). This is not a popular choice
in cosmology since $r$ will become multi-valued if the null cone
starts to refocus at the AH, a situation that can be avoided by
making use of $y$ as a distance on the null cone in terms of an
affine parameter $\nu$. However, it has been argued in
\cite{hellaby06} that the position of the AH is an observable
property by itself. Knowing this, the model can therefore be adapted
to compensate for refocussing possibly in a similar approach to that
used in \cite{lu07}. The details of such a development for the null
cone will however not be considered in the current paper and the
calculations will be limited to that region of the PNC interior to
the refocussing horizon.  A second aspect of $r$ is that it is not
comoving \emph{viz.} $v^1 \neq 0$. The third difference is that the
direction of the null cone is to the future. Changing the direction
of the null cone is straightforward by recognising that on a radial
null geodesic ($ds = d\theta = d\varphi=0$ and $du \neq 0$) from the
metric (\ref{sec:civp-met}):
\begin{align}
\int^r _{r_1} dr = -\frac{1}{2}\int^u _{u_1}{\left(1+\frac{W}{r}
\right) du} = \frac{1}{2}\int^{u_1} _{u}{\left(1+\frac{W}{r}
\right)du} .
\end{align}
Therefore changing the direction of numerical integration can be
used to determine the PNC instead of a future null cone.

\subsection{Cosmological observations}

The observational properties on the initial PNC as derived by
Bishop and Haines \cite{bish96} will be presented in a modified form
to separate the redshift $z$ and the diameter distance $r_0$.
The redshift in terms of the luminosity distance
($d_L$) is taken to be directly observable to a sufficiently
accurate order in the region where the null cone CIVP is well
behaved. The reciprocity theorem is used to find a relation between
$z$ and $r_0$ (see \cite{ellis71})
\begin{align}
z(d_L) \textrm{~~and~~} d_L = (1+z)^2 r_0 \Rightarrow \hat{z} =
z(r_0).
\end{align}
The redshift is directly related to the time component of the
contravariant velocity:
\begin{align}
1+\hat{z} = v^0
\end{align}
which can be used to determine the covariant velocity $v_1$:
\begin{align}
v_1(u_0,r) = - e^{2 \beta} v^0 = - e^{2 \beta} (1+\hat{z}) .
\label{sec:civp-v10}
\end{align}

The observed number count ($n$) is a directly measurable function of
$z$ and to be used as input to the CIVP once it
has been related to the proper density $\rho$. Since $\hat{z}(r_0)$ is
known, $n(z)$ can be rewritten as $\hat{n}(r_0) = n(\hat{z}(r_0))$.
It was shown in \cite{bish96} that the proper number count can be
written as:
\begin{align}
N = \frac{n}{(1+z)} e^{-2\beta} .
\end{align}
In terms of the diameter distance, the proper number count can be
written as:
\begin{align}
\hat{N}(r_0) = \frac{\hat{n}}{(1+\hat{z})}e^{-2\beta} .
\end{align}
The proper density is then related to the proper number count:
\begin{align}
\rho(u_0,r_0) = f(\hat{N}). \label{sec:civp-rho0}
\end{align}
The details of this relation will not be considered at this stage
but in principle it must take into account aspects such as dark
matter and source evolution, preferably with factors independent of
an already assumed cosmological model.

Solving the CIVP from observational quantities, as with the observer
coordinate method, requires that the values on the initial null cone
be determined from observational quantities. These values will then
be used as the initial values for the evolution into the local past
null cone. The solution on the initial null cone follows from
equations (\ref{sec:civp-v10}) and (\ref{sec:civp-rho0}) but they
are dependent on $\beta$ which is still unknown. Using equations
(\ref{sec:civp-beta}), (\ref{sec:civp-v10}) and
(\ref{sec:civp-rho0}), a differential equation for $\beta$ which is
only dependent on observational quantities for the initial null cone
can be set up as:
\begin{align}
\beta_{,r} &= 2\pi r f(\hat{N}) (- e^{2 \beta} (1+\hat{z}))^{2} .
\end{align}
This equation can be solved numerically with standard ODE techniques
provided that $f(\hat{N})$ is well behaved.

\section{Numerical implementation}
\label{s-Num}
\subsection{Code outline}

The code described in this section is based on the general 3D code
developed in \cite{bish97} and \cite{bish99} but significantly
simplified for spherical symmetry. The numerical scheme applied is a
combination of second order finite difference methods based on steps
half way between the $r$ and $u$ grid points on a regular grid. The
objective has been to obtain overall second order convergence in
both space and time up to a distance reasonably close to the
location where refocussing will occur.

The procedure described here, is to solve equations
(\ref{sec:civp-beta}-\ref{sec:civp-v0}) on a rectangular grid, which
is described in more detail in section \ref{sec:code-grid}. Solving
the hypersurface equations is done with a central difference method
on half steps between the $r$-grid points, using:
\begin{align}
g_{j}^i & = g_{j-1}^i + \frac{\Delta r}{2}(g_{,r \;j}^i +
g_{,r\;j-1}^i)
\end{align}
with $i$ being the time step and $j$ the radial step. Here
$g_{,r\;j}^i$ is calculated by substituting known values into
equations (\ref{sec:civp-beta}) and (\ref{sec:civp-W}). In order to
solve the evolution equations, (\ref{sec:civp-v1}) and
(\ref{sec:civp-rho}), their general form is notated as:
\begin{align}
v_{1,u} = F_{v1} \textrm{~~and~~} \rho_{,u} = F_{\rho}
\label{sec:code-gu}
\end{align}
and as explicit finite differences on a time half step they are
written as:
\begin{align}
v_{1j}  ^{n+1} = v_{1j} ^{n} + \Delta u F_{v1\;j} ^{n+1/2}
\textrm{~~and~~} \rho_{j}  ^{n+1} = \rho_{j} ^{n} + \Delta u F_{\rho
j} ^{n+1/2} .
\end{align}
Here, $n$ is a time iterator that will approach $i$. In these
equations, the numerical values at the point $(i,j)$ are used to
evaluate the matter terms while the hypersurface derivatives are
obtained by substituting the point values directly into the
numerical forms of equations (\ref{sec:civp-beta}) and
(\ref{sec:civp-W}). Radial matter derivatives are calculated making
use of standard central difference formulae (see for instance
\cite{burden93} p.160-161).

After setting up a suitable grid (as described in section
\ref{sec:code-grid}), the numerical algorithm can be summarised in
the following steps:
\begin{itemize}
\item[i.] Set the $\rho$ and $v_1$ initial values on to the initial grid
points. These values will, in principle, be obtained from
observations.
\item[ii.] Calculate $\beta$, $W$ and $v_0$ from $\rho$ and $v_1$ on the initial null cone.
\item[iii.] Calculate $F_{j} ^1$ the from the values of $v_1$, $\rho$
, $\beta$, $W$ and $v_0$.
\item[iv.] Set $F_{j} ^{n+1/2} = F_{j} ^1$ and calculate $v_1$ and
$\rho$ as an initial approximation that will approach the actual
values with subsequent iterations.
\item[v.] Use the new values of $v_1$ and $\rho$ to calculate
$\beta$, $W$ and $v_0$ and their radial derivatives.
\item[vi.] Calculate $F_{j}^{n+1/2}$ from values in (v) and again $v_1$ and
$\rho$.
\item[vii.] Test the calculations in vi for accuracy and convergence. If they are
sufficiently accurate, move to the next time step, otherwise repeat
steps v. and vi. with the new values of $v_1$ and $\rho$.
\end{itemize}

\subsection{Grid and domain of calculation} \label{sec:code-grid}

\begin{figure}[htb!]
\centerline{
    \includegraphics[height=0.35\textwidth]{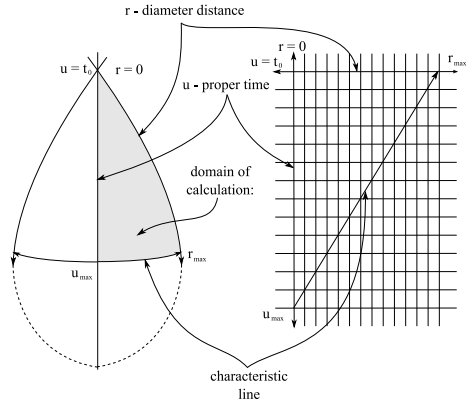}
} \caption{The past null cone calculated on a rectangular grid}
\label{code-grid}
\end{figure}

The domain of calculation is the interior region of the past null
cone starting from the present space-time location up to a region
approaching the point where the null cone starts to refocus. A
rectangular grid is used to represent the null cone as illustrated
in figure \ref{code-grid}. An important aspect of the code is that,
besides the coordinate conditions $\beta(u,0)=W(u,0)=0$, the
evolution scheme is unconstrained on the inner and outer radial
boundary i.e. no central or outer conditions will be required on the
$r$ coordinates. These boundaries do, however, still require to be
treated differently from the interior region.

On the central world line ($r=0$) in figure \ref{code-grid} and its
close vicinity, the CIVP equations are not
suitable in general form (equations (\ref{sec:civp-v1}) and
(\ref{sec:civp-rho})), which is evident from the occurrence of $r$
denominators. Consequently, the $r\approx 0$ region is calculated by
making use of series expansions. In order to do this, all the
variables in the CIVP equations are replaced by power series
expressions from which the terms can then be calculated from the
resulting coefficients. In the expansions, only the terms required
to produce second order accuracy will be used, for instance for the
density evolution:
\begin{align}
\rho_{,u} &= \rho_{u0} + \rho_{u1} r + \rho_{u2}r^2 +
\mathcal{O}(r^3) = F_{\rho} \label{sec:series-rhou} \;.
\end{align}
The region where the series solution meets with the CIVP solution,
also requires special treatment to avoid artificial instabilities
which has been done by smoothing out the merger region with a
weighted average between the two solutions.

\begin{figure}[htb!] \centerline{
    \includegraphics[height=0.45\textwidth, angle=-90]{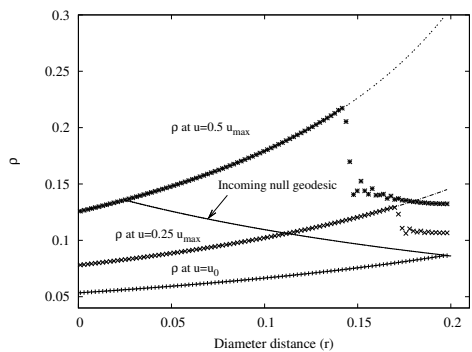}
}\caption{Unstable regions being excluded using a null geodesic on
the outer boundary. The figure is based on a portion of a normalised
Einstein-de Sitter calculation.} \label{civp-rho-rmax}
\end{figure}

In order to evolve the CIVP equations without any
boundary condition on the outer radial boundary, the outer boundary
must be an incoming null hypersurface. This is needed for well-posedness,
and if violated leads to an numerical instability.
An incoming
null ray (i.e. on a radial null geodesic $ds = d\theta = d\varphi=0$
and $du \neq 0$) is selected for the outer radial limit and is
defined from the metric in the following way:
\begin{align}
r_{outer} - r_{outer \;u_1} = -\frac{1}{2}\int^u
_{u_1}{\left(1+\frac{W}{r} \right) du}
\end{align}
Numerically this can be solved by using an iterative scheme which
starts with estimating $W=0$ and is then repeated until the values
of $W$ and $r$ converge. Since the value of $r$ does not necessarily
fall directly on grid points, $W$ is determined by linearly
interpolating between the closest grid points on a specific $u_i$
grid line. The behaviour of the density evolution is shown in figure
\ref{civp-rho-rmax}, where the computation extends beyond the null
ray cut-off region in order to illustrate the stability issue.

\section{Code verification}
\label{s-Codev}
\subsection{LTB models}

The spherically symmetric CIVP for cosmology is effectively the
Lema\^{i}tre-Tolman-Bondi (LTB) model in null coordinates.
Originating and developed by Lema\^{i}tre \cite{lem33}, Tolman
\cite{tol34} and Bondi \cite{bon47}, the LTB model is a
generalisation of the Friedmann-Lema\^{i}tre-Robertson-Walker (FLRW)
model with inhomogeneities allowed in the radial coordinate. The
parabolic case of the LTB model is of interest for the
investigations in this paper and will be used to verify the CIVP
code. The metric in comoving synchronous coordinates is:
\begin{align}
ds^2 = -dt^2
    + [R_{,r}(t,r)]^{2}dr^2 + [R(t,r)]^2\{d\theta ^2 + \sin ^2 \theta d\varphi
    ^2\}
    \label{sec:civp-ltbm}
\end{align}
Here, $t$ is the cosmic (proper) time, $r$ is a comoving radial
coordinate with $\theta$ and $\varphi$ the inclination and azimuth
angles. $R(t,r)$ is the areal radius and $4\pi R^2$ defines the
proper surface area of a sphere with coordinate radius $r$ at a
constant time slice \cite{ellis98}. Substituting into the Einstein
field equations and solving gives (see \cite{krasinski06}):
\begin{align}
R(t,r) = \left[\frac{9}{2} M(r) (t-t_{B}(r))^{2}\right]^{1/3}
\textrm{~~and~~} \rho(t,r) = \frac{M_{,r}}{4 \pi \; R^2 R_{,r}}
\textrm{ .} \label{sec:ltb-e0}
\end{align}
$M(r)$ is the \emph{active gravitational mass} which is the mass
contributing to the gravitational field and $t_B(r)$ is defined as
the \emph{bang time function}, which is a surface defined by the
local time at which $R=0$.

In order to provide data to verify the CIVP code against known
solutions, LTB models need to be transformed to null coordinates
which will then be used as input data on the initial null cone as
well as comparative data inside the null cone. Numerous LTB models
exist which include the FLRW models, other physical realistic models
and also many models with uncertain physical relevance. Many
physical LTB scenarios have been investigated that are possible
candidates of the actual universe (see for instance \cite{gar08})
which will all be interesting scenarios to run on a CIVP code.
However, for the purpose of code verification, it will be conducive
to work with models that are mathematically simple, well documented
and physically relevant. The class of LTB models that will be
investigated is referred to as \emph{bang time models} and makes use
of the bang time function, $t_B$, as a mechanism to induce
inhomogeneities into a universe. By selecting simple functions for
$t_B$, it has been demonstrated that inhomogeneities can mimic the
effect of inflation (solving the horizon problem) \cite{celerier98,
celerier00i}, describe the cosmic microwave background (CMB) dipole
anisotropy \cite{schneider98}, and how inhomogeneities can reproduce
the effect of supernovae-redshift dimming without the requirement of
dark energy \cite{celerier00}.

For these models, $M(r) = M_0=r^3$ is chosen as a coordinate
condition where $M_0$ is a constant which for illustrative purposes
is set to $2/9$. Equations (\ref{sec:ltb-e0}) then reduce to:
\begin{align}
R(t,r) &= r (t-t_{B}(r))^{2/3} \label{sec:civp-R}
\end{align}
with:
\begin{align}
R_{,r}(t,r) &= \frac{t-t_{B}(r) - 2/3 \,r\,
t_{B,r}(r)}{(t-t_B(r))^{1/3}} \label{sec:civp-Rx}
\end{align}
and $\rho$ in equation (\ref{sec:ltb-e0}) to:
\begin{align}
\rho(t,r) &= \frac{1}{2 \pi (t-t_B(r))(3t - 3t_B(r)-2rt_{B,r}(r))} .
\label{sec:civp-ltbrho}
\end{align}

By selecting $t_B(r)=0$, it can be seen that, equation
(\ref{sec:civp-ltbrho}) becomes the Einstein-de Sitter (EdS) model,
which has been used as a test compared in previous work
\cite{bish96}. If $t_B(r) \neq 0$ and $t_{B,r}(r)=0$, the time of
the initial singularity is adjusted and the age of the Universe
changes. For non-constant functions, the initial singularity becomes
a singular surface and the age of the Universe becomes subject to
the position of an observer (i.e. the age of the Universe depends on
$r$). Thus, a variety of models can be generated for testing the
code on parabolic spatial sections.

\subsection{Coordinate transformations}

In order to generate results from LTB models on the null cone, a
coordinate transformation must be made from equation
(\ref{sec:civp-met}) to equation (\ref{sec:civp-ltbm}). For
simplicity, the null cone metric will be written in the form:
\begin{align}
ds^2 = -h_u du^2 - 2h_rdudr + r^2 \{d \theta ^2 + \sin ^2 \theta d
\varphi ^2\} . \label{sec:civp-ncm}
\end{align}
and the coefficients will be rewritten in terms of $W$ and $\beta$
afterwards. Also, LTB variables that have the same symbols as the
null cone variables will be notated using a tilde e.g.
$r_{LTB}=\tilde{r}$, $v_{1\;LTB}=\tilde{v}_1$, $g_{LTB}=\tilde{g}$
etc.

The essence of the transformations is to relate the $(u,r)$ coordinates to
corresponding $(t,\tilde{r})$ values. An expression for $r$ in terms
of $\tilde{r}$ and $t$ can be obtained by comparing the coefficient
of $(d\theta ^2 + \sin ^2 \theta d \varphi)$
in the two coordinate systems, leading to
\begin{align}
R(t,\tilde{r}) = r  \label{sec:rxt}
\end{align}
A second expression containing $u$,
$\tilde{r}$ and $t$ can be found by making a transformation
from LTB coordinates to find in null cone coordinates the term
$g^{00}$, which is equal to zero. A first order partial differential
equation follows:
\begin{align}
0 = \frac{\partial u}{\partial \tilde{r}}
            - [R_{,\tilde{r}}(t,\tilde{r})] \frac{\partial u}{\partial t}
            \textrm{~~with~~} u(t,0) = t \label{sec:two}
\end{align}
which, using the method of characteristics, can be written as:
\begin{align}
\frac{dt}{d\tilde{r}} = - [R_{,\tilde{r}}(t,\tilde{r})].
\label{sec:char}
\end{align}
Depending on the complexity of the $R_{,\tilde{r}}$ term, either
numerical or analytic methods can be used to solve this equation.

Expressions containing partial derivatives for the other null cone
terms can be obtained from covariant transformations. For $h_r$ and
$h_u$ these are:
\begin{align}
h_r &=
    \frac{\partial t}{\partial u}
    \frac{\partial t}{\partial r}
    - [R_{,\tilde{r}}(t,\tilde{r})] ^2
    \frac{\partial \tilde{r}}{\partial u}
    \frac{\partial \tilde{r}}{\partial r} \label{sec:tran-hr} \\
h_u &=
    \left( \frac{\partial t}{\partial u} \right) ^2
    - [R_{,\tilde{r}}(t,\tilde{r})] ^2
    \left(\frac{\partial \tilde{r}}{\partial u}\right) ^2 \label{sec:tran-hu}
\end{align}
Having values for $h_u$ and $h_r$, the Bondi-Sachs coefficients can
be obtained from:
\begin{align*}
\beta = \frac{1}{2} \ln{\mid h_r \mid} \textrm{~~and~~} W = r(h_u
h_r - 1)
\end{align*}

In order to write $v_1$ in null coordinates, requires the
transformation of the comoving velocity ($\tilde{v}^{\mu} =
(1,0,0,0)$ and $\tilde{v}_{\mu} = (-1,0,0,0)$) into null
coordinates.
\begin{align}
v^0 = \frac{\partial u}{\partial t} \textrm{~~,~~} v^1 =
\frac{\partial r}{\partial t} \textrm{~~,~~} v_0 = -\frac{\partial
t}{\partial u} \textrm{~~and~~} v_1 =-\frac{\partial t}{\partial
r}\label{sec:tran-v1}
\end{align}
The five point difference equations in \cite{burden93} were used to
determine the grid values of equations (\ref{sec:tran-hr}),
(\ref{sec:tran-hu}) and (\ref{sec:tran-v1}); $v_1$ from
(\ref{sec:tran-v1}) together with $\rho$ from
(\ref{sec:civp-ltbrho}) are used as the input data on the initial
null cone.

\subsection{Verification results}

A simple choice of bang function is implemented as a verification
model:
\begin{align}
t_B(r) = b r , \label{sec:civp-tB}
\end{align}
with $b$ being a constant. This simplifies equations
(\ref{sec:civp-R}), (\ref{sec:civp-Rx}) and (\ref{sec:civp-ltbrho})
to:
\begin{align}
R(t,r) &= r (t-b r)^{2/3} \label{sec:civp-Rkx} \\
R_{,r}(t,r) &= -\frac{1}{3} \frac{(-3 t + 5 b r)}{(t-b r)^{1/3}}
\label{sec:civp-Rxkx} \\
\rho(t,r) &= \frac{1}{2 \pi (t-br)(3t -5br))} .
\label{sec:civp-ltbrhok}
\end{align}

\begin{figure}[htb!]
\centerline{
    \includegraphics[height=0.5\textwidth, angle=-90]{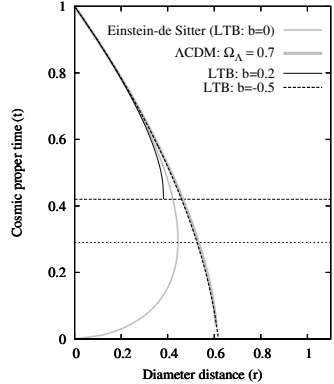}
} \caption{The past null cone of different bang time models compared
to the Einstein-de Sitter model and the $\Lambda$CDM model.}
\label{sec:pnc-ltb}
\end{figure}

The effect of different values of $b$ on the shape of the PNC is
shown in figure \ref{sec:pnc-ltb} where different models have been
scaled and transposed relative to an observer located at the vertex
of the PNC of a normalised EdS universe. The value $b=0$ is exactly
the EdS model and $b>0$ shifts the age of a universe to a younger
age as $r$ increases while $b<0$ provides the opposite effect where
a universe is shifted to an older age as $r$ increases. The latter
case is particularly interesting since it provides a mechanism to
mimic a cosmological constant \cite{celerier00}. In figure
\ref{sec:pnc-ltb} a reasonable match on the shape of the
$\Lambda$CDM ($\Omega_{\Lambda}=0.7$) PNC is shown by a $b=-0.5$
bang time model before the PNC refocusses. This is however not an
exact physical match but provides a useful verification model since
the shape of the null cone is a critical aspect on the stability of
a CIVP code.

The test cases presented in this section are the $b=0$ (EdS) case up
to $0.82 \;u_{max}$ and $0.82 \;r_{max}$ and the $b=-0.5$ case up to
$0.75 \;u_{max}$ and $0.8 \;r_{max}$ with $u_{max}$ and $r_{max}$
the time and distance of the AH. The EdS case provides a useful
baseline since its solution is exactly known in null coordinates
while the $b=-0.5$ case is representative of an LTB model with a
$\Lambda$CDM-like PNC. It should be noted that the EdS model is {\bf
not} homeogeneous with respect to the null cone radial coordinate
$r$ and is therefore also representative of more general LTB models.
The limits chosen are the maximum values where stable solutions were
achieved. Extending these limits further causes the code breaks down
rapidly, which is the expected behaviour close to the AH.

The radial outer limits of the initial PNCs correspond to
observations at $z=0.47$ and $z=0.48$ in the EdS and LTB cases
respectively. By reducing the extent of $u$ limit, the radial extent
can be increased and conversely reducing the radial limit allows the
$u$ limit to be increased, for instance values of $0.2 \;u_{max}$
and $0.99 \;r_{max}$ ($z=1.02$) also produce stable results for the
EdS model. The stability in terms of the radial and evolution limits
is dependent on the input data and the simulations in section
\ref{s-LTB} were stable up to $z\approx1$, which is a significant
region for supernovae redshift-distance observations. In this
section values were chosen that were useful for demonstrating
accuracy and stability on normalised models.

\subsubsection{Physical variables}

\begin{figure}[h!]
\begin{center}$
\begin{array}{ll}
\hspace{-30pt}
\includegraphics[width=0.35\textwidth, angle=-90]{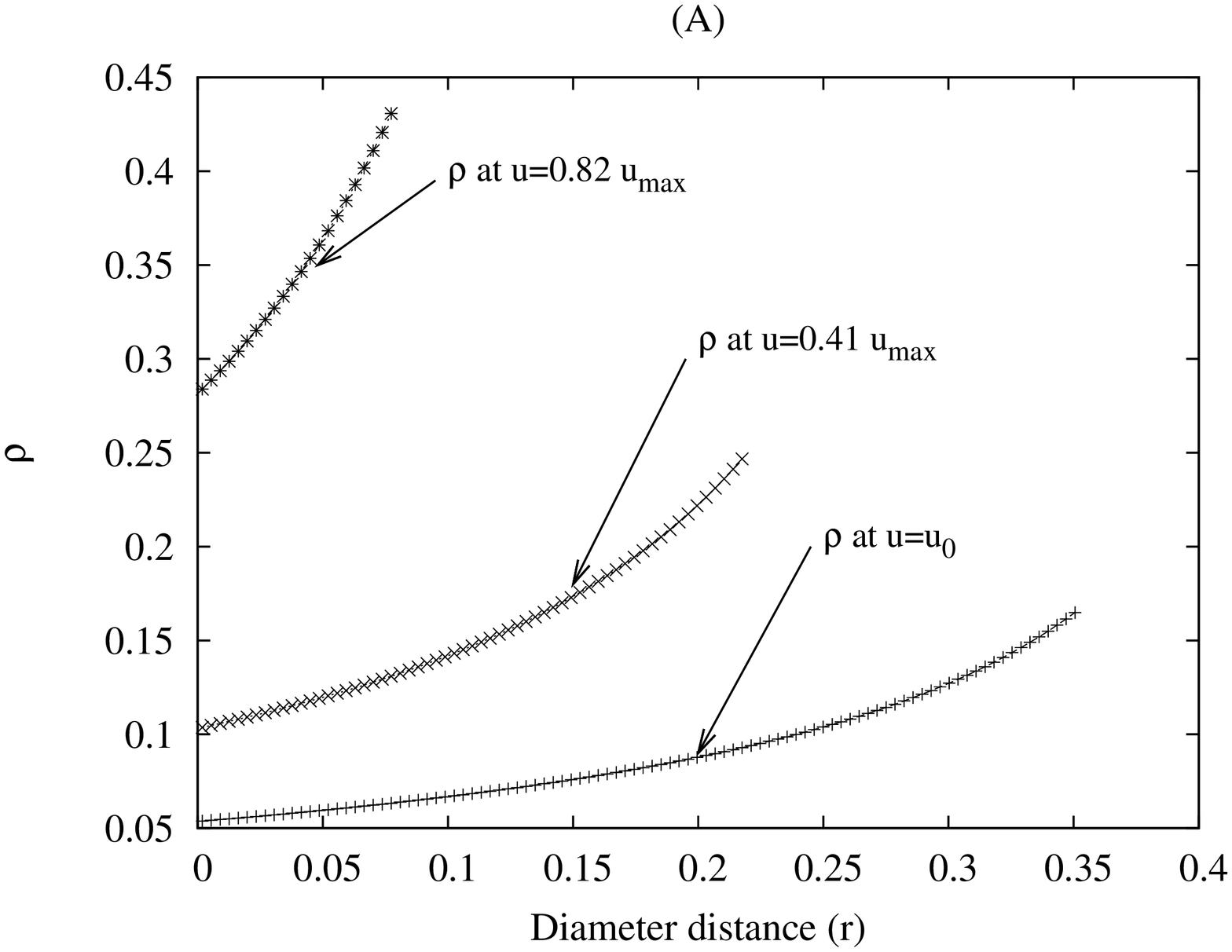} &
\hspace{-30pt}\includegraphics[width=0.35\textwidth,
angle=-90]{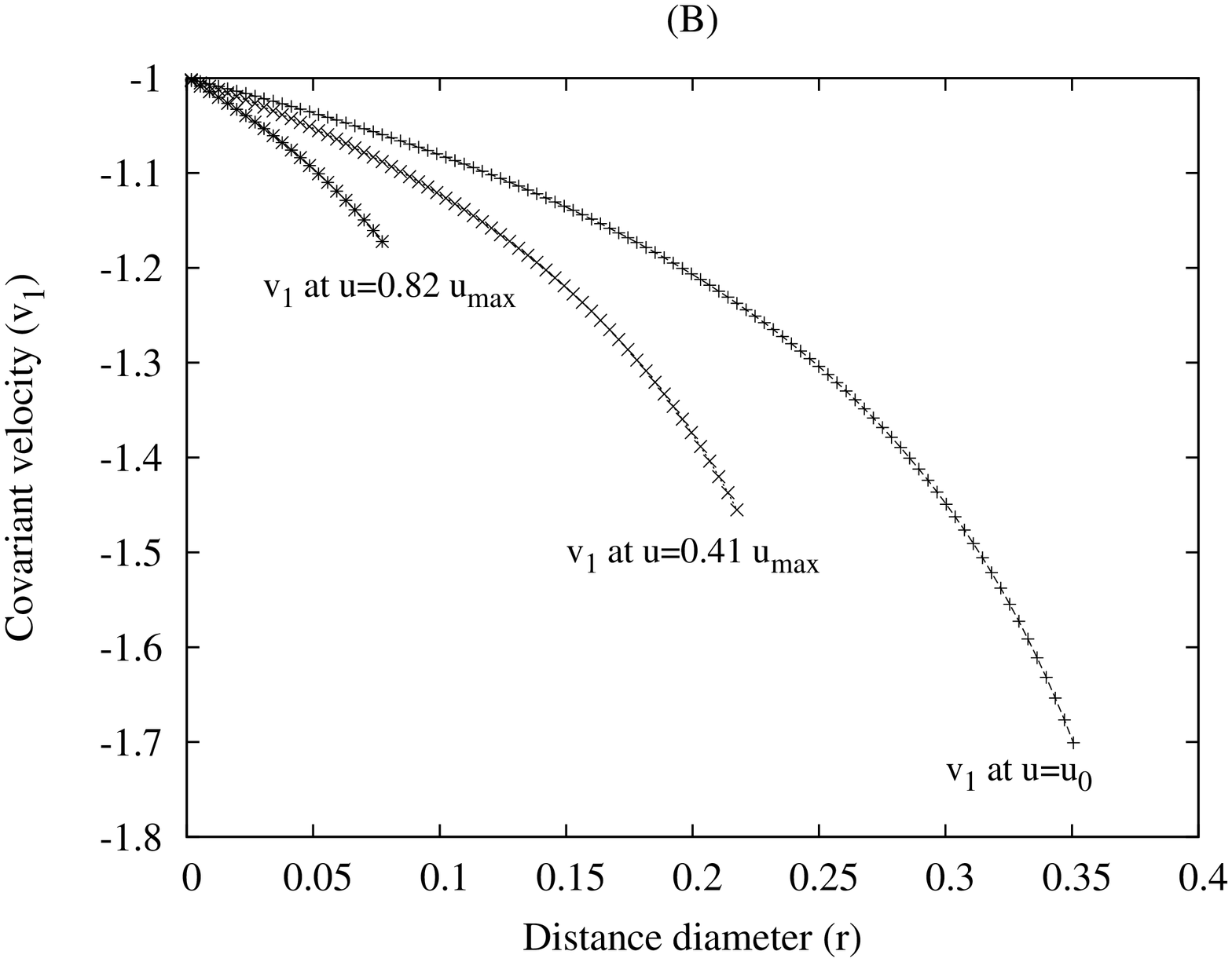} \vspace{-10pt} \\ \hspace{-30pt}
\includegraphics[width=0.35\textwidth, angle=-90]{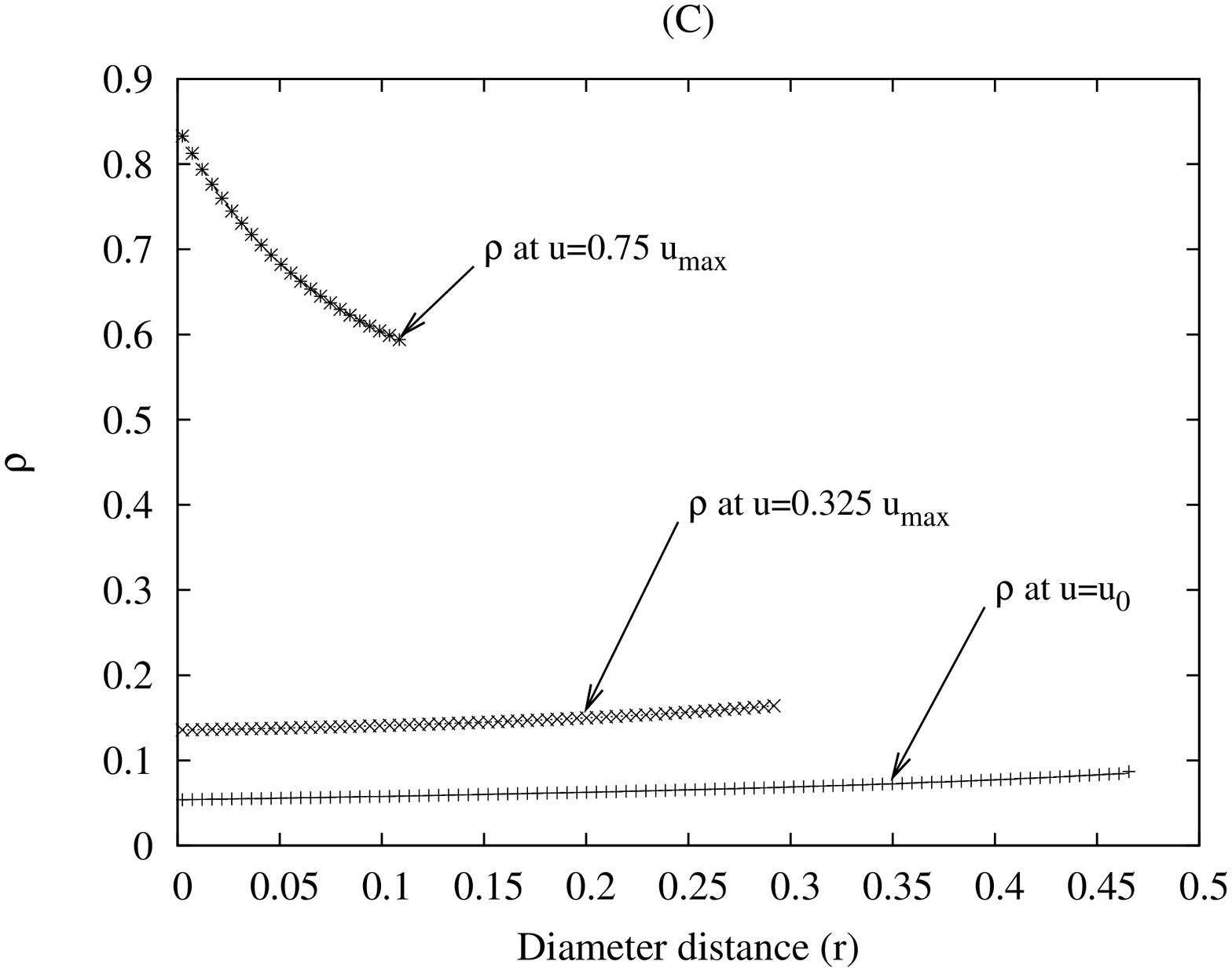} &
\hspace{-30pt}\includegraphics[width=0.35\textwidth,
angle=-90]{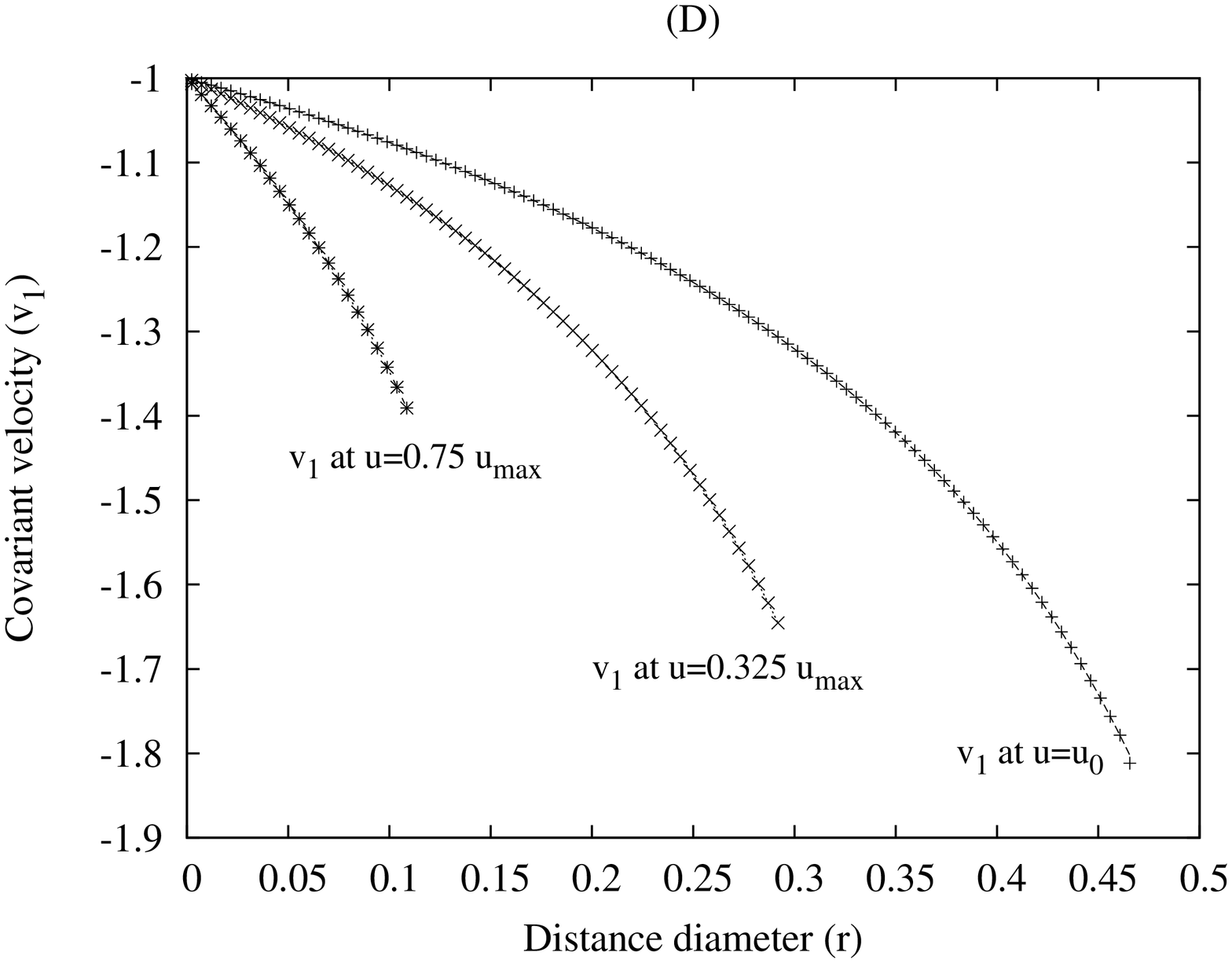}
\end{array}$
\end{center} \vspace{-10pt} \caption{$\rho$ and $v_1$ against
$r$ on null cones in the past for the Einstein-de Sitter model (A)
and (B) and the $b=-0.5$ bang time model (C) and (D).}
\label{sec:res-rho-v1}
\end{figure}

Figure \ref{sec:res-rho-v1} shows the density ($\rho$) and covariant
velocity ($v_1$) of the two models. The plots contain lines, which
represent the exact values, and points representing the code values.
The exact and code values are close enough not to show visible
deviations. The density distributions of the two models on the
oldest PNC are interesting to compare where the $b=-0.5$ model
clearly displays the effect of the bang time surface.

\subsubsection{Local error propagation and convergence}

\begin{figure}[h!]
\vspace{-20pt}
\begin{center}$
\begin{array}{ll}
\hspace{-30pt}
\includegraphics[width=0.35\textwidth, angle=-90]{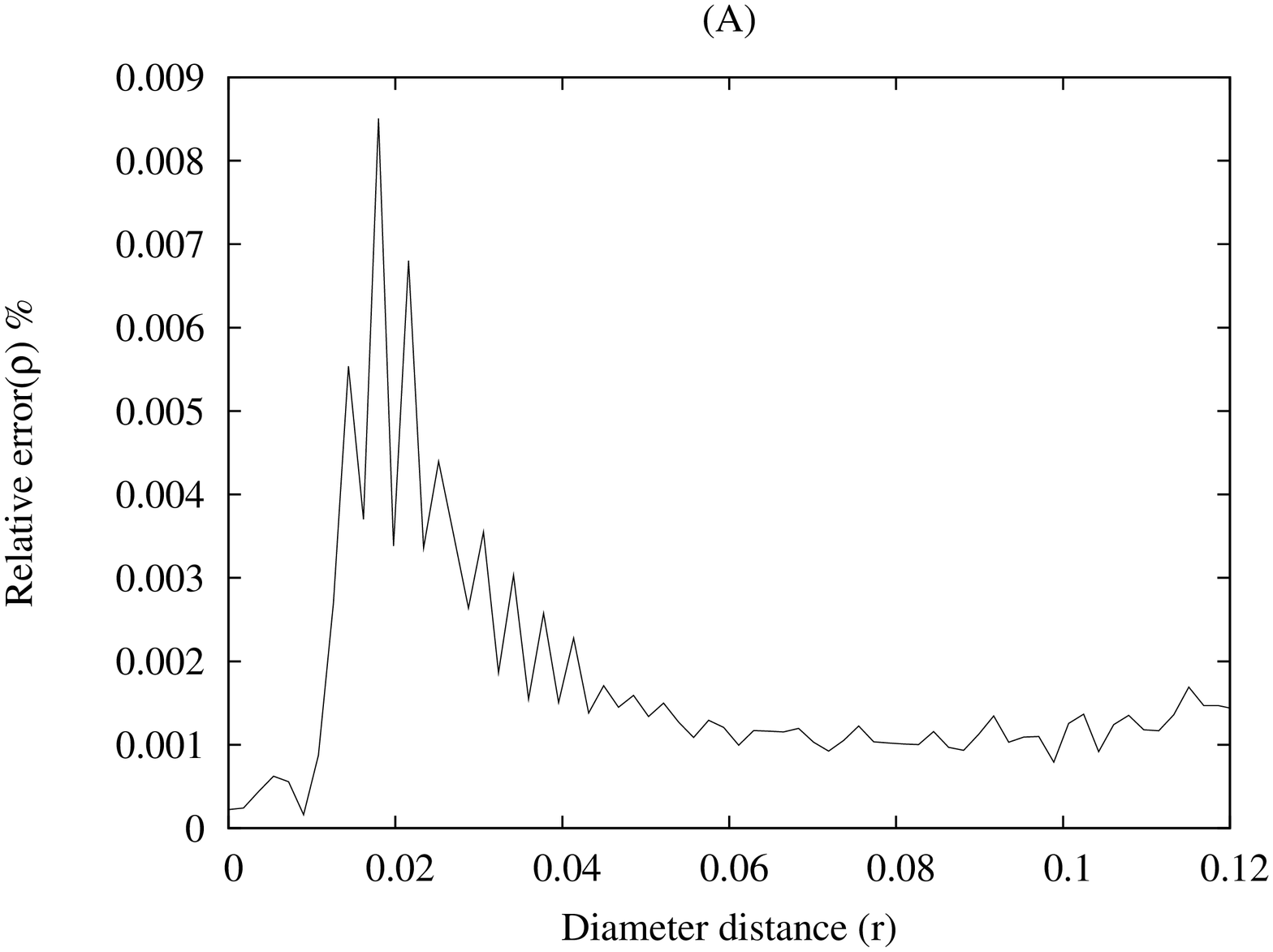} &
\hspace{-30pt}\includegraphics[width=0.35\textwidth,
angle=-90]{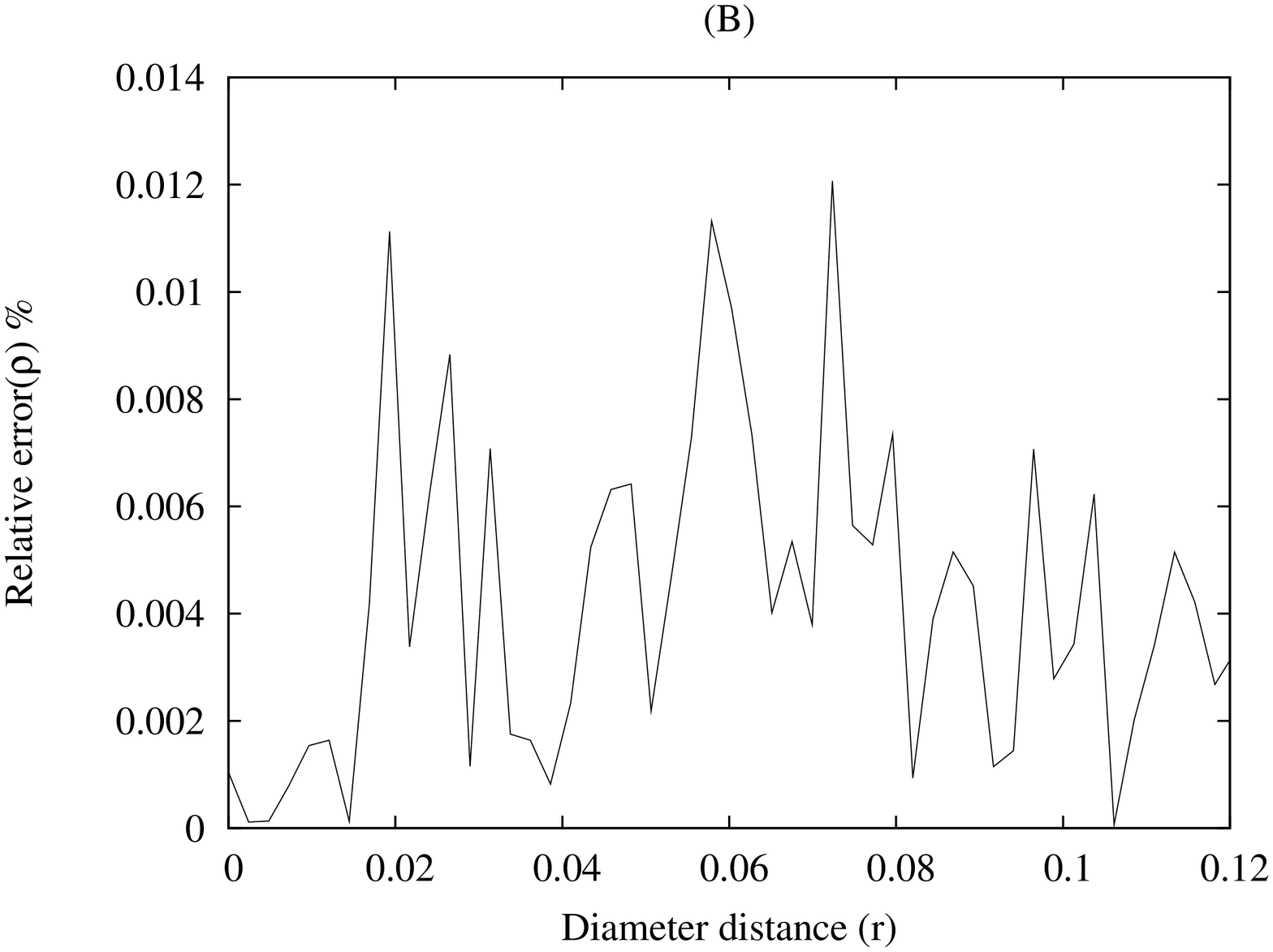}
\end{array}$
\end{center} \vspace{-10pt} \caption{Spatial error vs. $r$ for 200 $r$ $\times$ 1000 $u$ grid
points for the Einstein-de Sitter model (A) and the $b=-0.5$ bang
time model (B).} \label{sec:res-eds-ltb-error}
\end{figure}

Figure \ref{sec:res-eds-ltb-error} shows the radial distributions of
the relative errors on the oldest null cones. As a measure of
errors, the relative percentage of the difference between the exact
and code of the density calculation has been chosen as a
representative quantity. This choice has been based on different
test cases where small errors in the other variables, $v_1$, $W$ and
$\beta$ have become pronounced in the density calculation. The EdS
model having more accurate input values has a better local accuracy
but the series-CIVP merger region is more visible in the $r \approx
0$ region.

\begin{figure}[h!]
\begin{center}$
\begin{array}{ll}
\hspace{-30pt}
\includegraphics[width=0.35\textwidth, angle=-90]{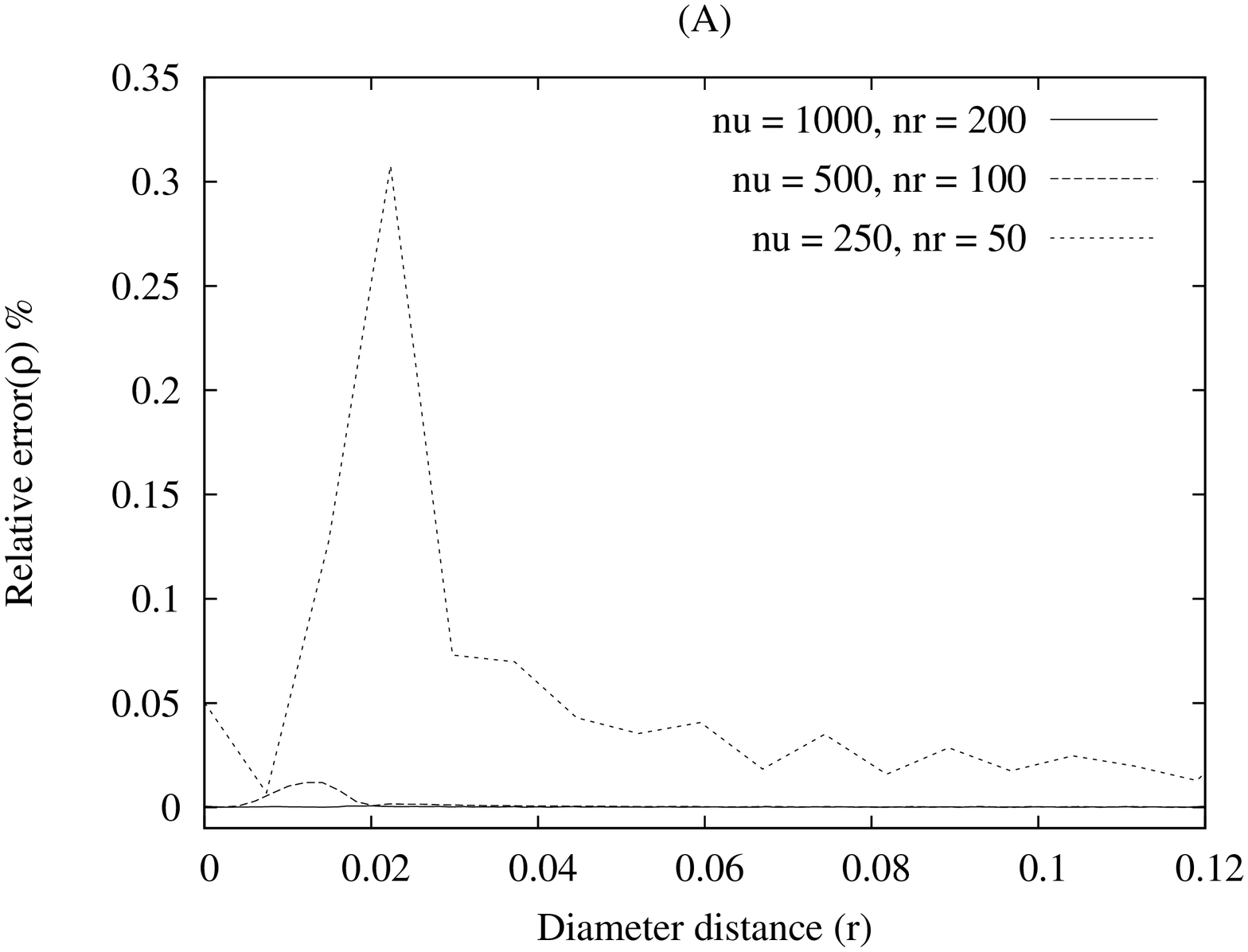} &
\hspace{-30pt}\includegraphics[width=0.35\textwidth,
angle=-90]{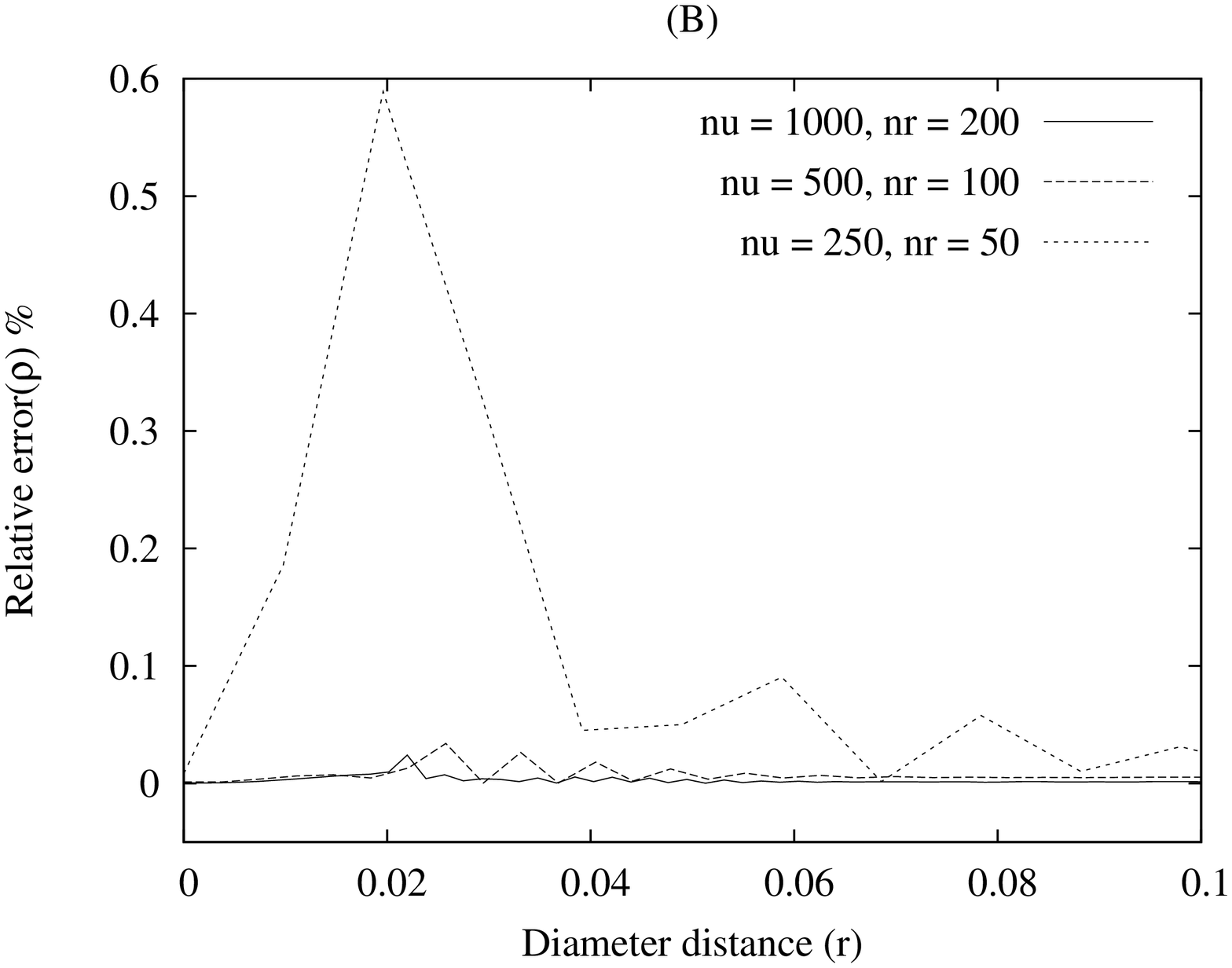}
\end{array}$
\end{center} \vspace{-10pt} \caption{Grid convergence: The error propagation for $u \times r$
grid resolutions of $1000\times 200$, $500\times100$ and
$250\times50$ for the Einstein-de Sitter model (A) and the $b=-0.5$
bang time model (B).} \label{sec:res-eds-ltb-conv}
\end{figure}

In figure \ref{sec:res-eds-ltb-conv}, errors are shown where grid
resolution has been refined three times. It can be seen that errors
in the region where the series solution is merged with the CIVP
solution is significantly larger at low grid resolutions but
converge rapidly to containable levels. The merging region between
the series and CIVP solution is particularly sensitive which
suggests that further refinement of this region will be required
especially if realistic observational data will be used as input in
future research.

\begin{figure}[h!]
\begin{center}$
\begin{array}{ll}
\hspace{-40pt}
\includegraphics[width=0.35\textwidth, angle=-90]{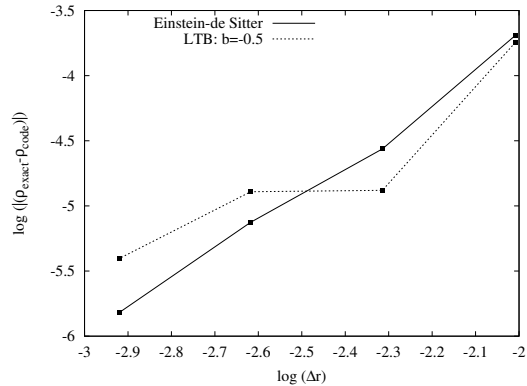}
\end{array}$
\end{center} \vspace{-10pt} \caption{Grid convergence: The slope of the $\log (error_{max})$
vs. $\log (\Delta r)$ line is $1.9$ for a $\Delta r / \Delta u
\approx 3.1$ grid for the Einstein-de Sitter model and $1.8$ for a
$\Delta r / \Delta u \approx 3.2$ grid for the $b=-0.5$ bang time
model.} \label{sec:res-convergence}
\end{figure}

In figure \ref{sec:res-convergence}, the convergence against the
radial grid size is shown. Here the slope of the line on a log scale
is $1.9$ and $1.8$ for the EdS and $b=-0.5$ cases respectively,
indicating that convergence approaches second order. The fact that
the code breaks down after being close to second order accurate is
as an accuracy consideration a useful feature since it doesn't
produce false results; it either works accurately or provides no
results.

\section{LTB vs. $\Lambda$CDM}
\label{s-LTB}
\subsection{Overview}

Following from the \emph{Isotropic Observations Theorems} in
\cite{musta98} (also see \cite{ellis98}), any reasonable set of
spherical symmetrical observations can be accommodated by some
distribution of inhomogeneities in an LTB model. Soon after the
discovery of dark energy from Supernovae Ia observations in 1998
\cite{perl99, ries98}, using this principle, it was demonstrated by
C\'{e}l\'{e}rier \cite{celerier00}, with others to follow (e.g.
\cite{tomita00}), that the same observational effects can be caused
by radial inhomogeneities without any need of a cosmological
constant. The concept of inhomogeneities mimicking dark energy has
been investigated many times since then, mostly as toy models
demonstrating the ambiguity of observations commonly accepted in
conventional cosmology. The main argument against LTB universes is
that they place the observer in the centre of the Universe which,
although not impossible is not a philosophically appealing idea.
Even so, with a CIVP code at hand, an interesting numerical
experiment is to test the historical evolution of an LTB model, with
a zero cosmological constant, when observational data representing
that of the $\Lambda$CDM model is interpreted as being caused by
inhomogeneities.

A simple simulation set up for this experiment is done by
calculating the density $\rho$ and covariant velocity $v_1$ on the
past null cone by transforming exact solutions for a flat
$\Lambda$CDM model onto the past null cone. These are then used as
input to the code and compared to the transformed model. In terms of
the LTB metric (\ref{sec:civp-ltbm}), using cosmological properties
at the current epoch ($t_0$), the solution of the flat $\Lambda$CDM
model is (see \cite{lidsey09}):
\begin{align}
R(t,\tilde{r}) = S(t)\,\tilde{r} = \left(
\frac{\Omega_{m0}}{\Omega_{\Lambda0}}\right)^{1/3} \left(\sinh\left[
\frac{3}{2} H_0 \sqrt{\Omega_{\Lambda0}} \, t\right] \right)^{2/3}
\tilde{r}
\end{align}
with the age of the Universe:
\begin{align}
t_0 = \frac{2}{3} \left(H_{0} \sqrt{\Omega_{\Lambda0}}\right)^{-1}
\sinh^{-1}\left[
\left(\frac{\Omega_{\Lambda0}}{\Omega_{m0}}\right)^{1/2} \right]
\end{align}
and the density distribution
\begin{align}
\rho_m = \frac{3 H_0^2}{8 \pi G}  \frac{\Omega_{m0} S_0^3}{S^3}.
\end{align}
Here, $\Omega_{m0}$ and $\Omega_{\Lambda0}$ are the current density
parameters for Baryonic matter and the cosmological constant
respectively and $H_0$ is the current Hubble constant. The values used
to represent the actual Universe were: $\Omega_{m0}=0.3$,
$\Omega_{\Lambda0}=0.7$ and $H_0=72$ Mpc s$^{-1}$ km$^{-1}$. An
important issue to notice here is that $\rho_m$ in the $\Lambda$CDM
model is determined by parameters related to the expansion and the
density content and not by an independent measure of matter
distribution such as number counts. The additional degree of freedom
introduced in the LTB model is therefore not purely satisfied by an
additional boundary condition which is a limitation that should be
borne in mind when interpreting redshift dimming as an LTB model.

\begin{figure}[h!]
\begin{center}$
\begin{array}{ll}
\hspace{-30pt}
\includegraphics[width=0.35\textwidth, angle=-90]{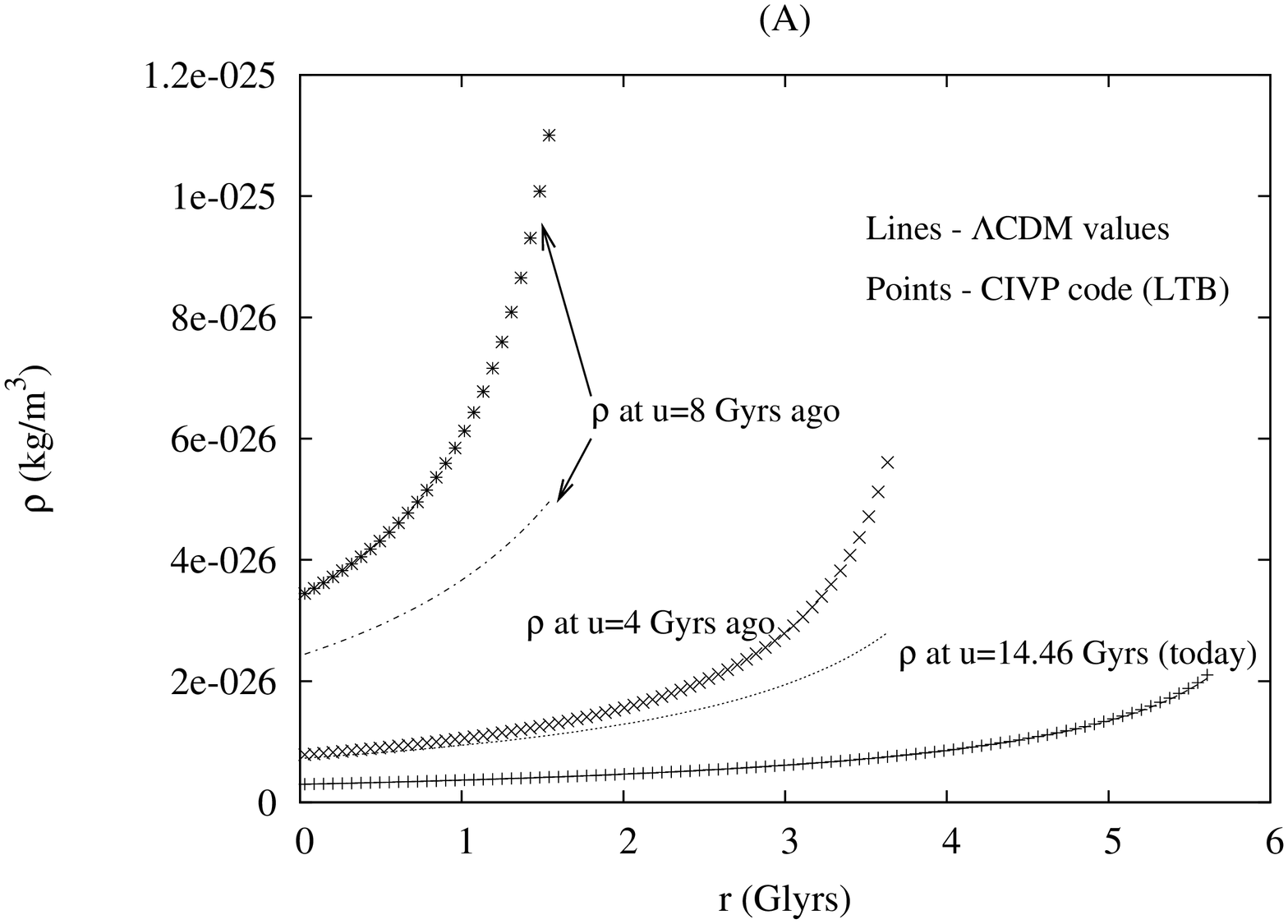} &
\hspace{-20pt}\includegraphics[width=0.35\textwidth,
angle=-90]{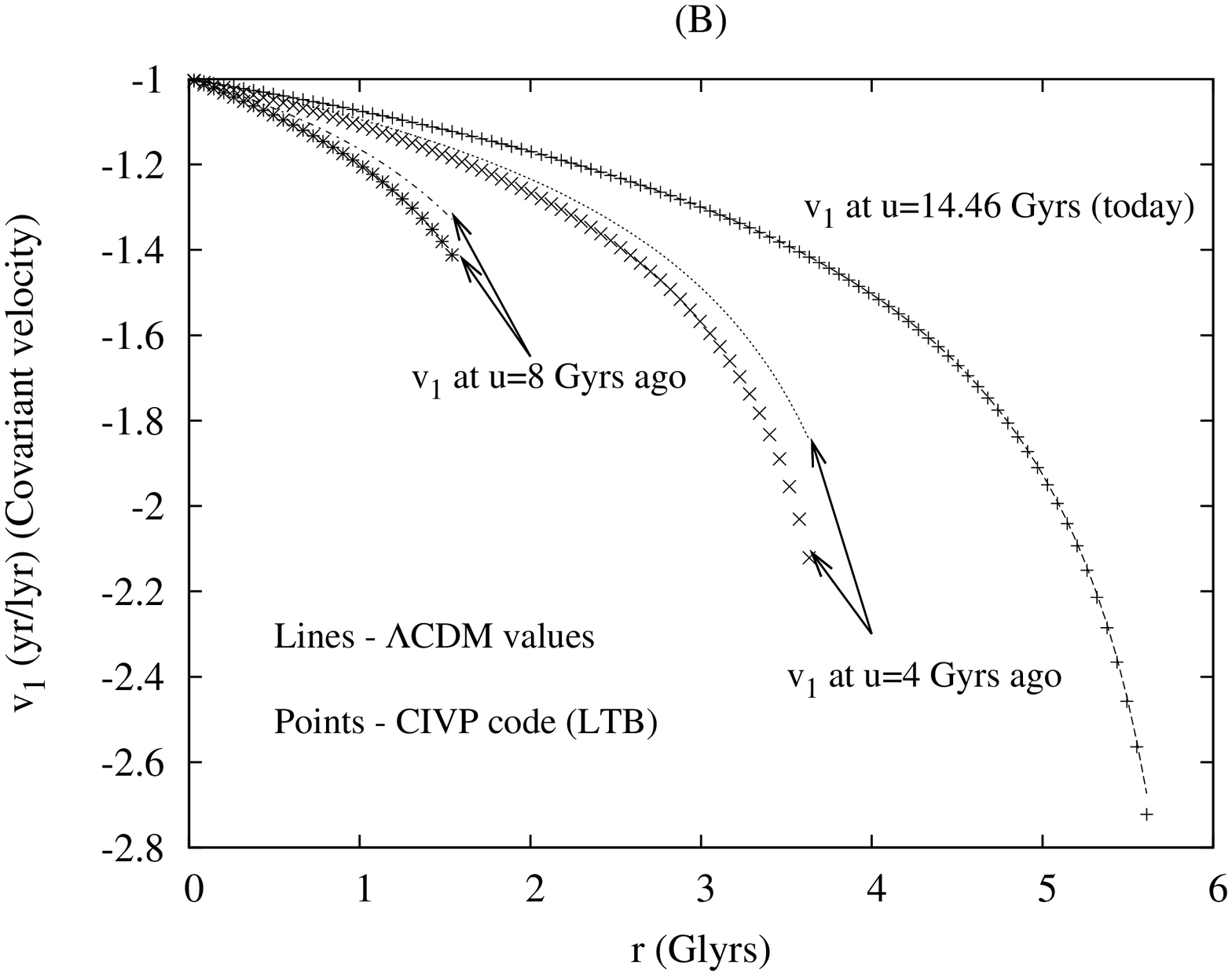}
\end{array}$
\end{center} \vspace{-10pt} \caption{Density distribution (A) and covariant velocity
(B) on past null cones at different proper times ($u$)}
\label{sec:cc-results}
\end{figure}

Figure \ref{sec:cc-results} shows the resulting LTB vs. $\Lambda$CDM
evolutions back in time. While it might currently not be possible to
distinguish these models from one another based on observations on
the past null cone, in the past, these universes are distinctly
different. In particular, the LTB universe seems to be heading
towards a singularity much faster than the $\Lambda$CDM universe and
would therefore be significantly younger, if the trend is to
continue beyond the calculations.

\subsection{LTB vs. $\Lambda$CDM matching}

A question arising from the results presented in figure
\ref{sec:cc-results} is: if the LTB null cones in the past would
also represent $\Lambda$CDM flat space null cones? In this section
an attempt is made to find a matching $\Lambda$CDM model for the
$t_0-8$Gyrs LTB null cone. The procedure followed here is to firstly
find a null cone in the past of a selected $\Lambda$CDM model for
which the density corresponds to the LTB density at $r=0$. This
gives a density curve which has the same starting point but the
slope of the curve differs to that of the LTB model. The slope of
the curve is then matched by adjusting the Hubble rate ($H_0$) until
it approximately corresponds to the LTB curve. The covariant velocities
($v_1$ as a function of $r$) are then compared to see, qualitatively,
if the models can be regarded as the same.

\begin{figure}[h!]
\begin{center}\vspace{-20pt} $
\begin{array}{ll}
\hspace{-30pt}
\includegraphics[width=0.35\textwidth, angle=-90]{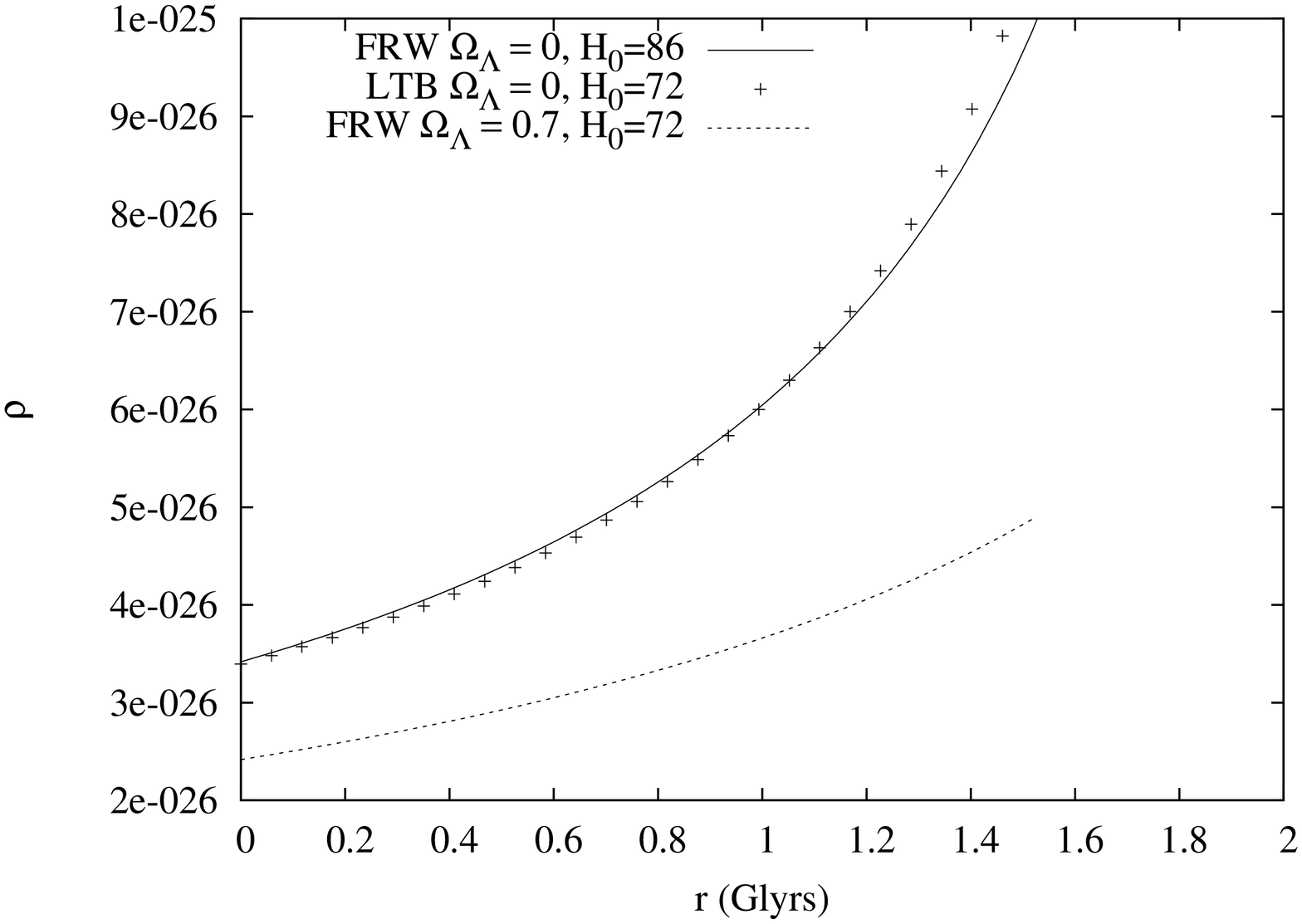} &
\hspace{-20pt}\includegraphics[width=0.35\textwidth,
angle=-90]{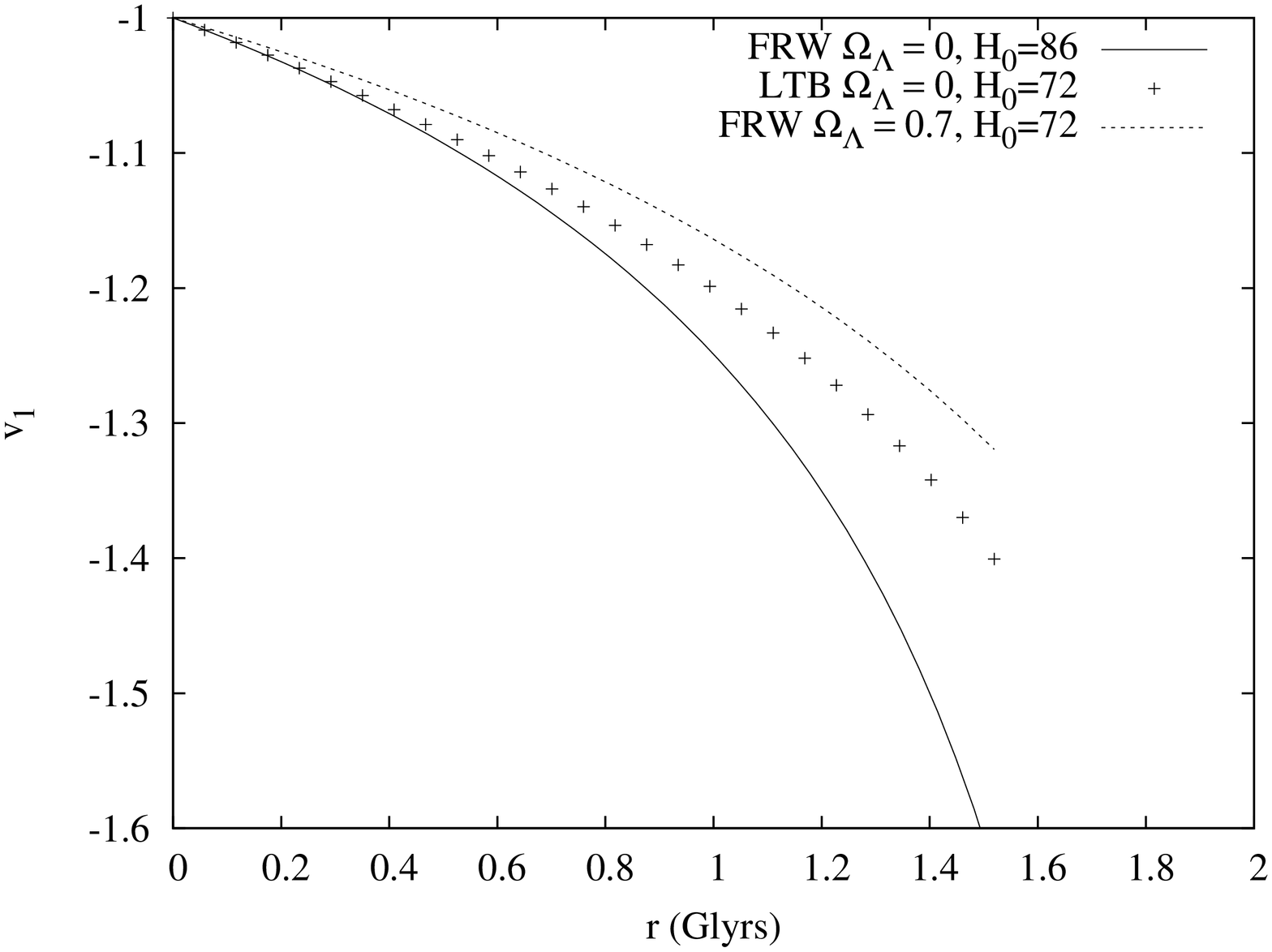}
\end{array}$
\end{center} \vspace{-20pt} \caption{Density distribution and covariant velocity
on the past null cones: LTB with $\Omega_\Lambda=0$ and FLRW with
$\Omega_\Lambda=0.7$ at $t_0-8$Gyrs and the best fit FLRW
$\Omega_\Lambda=0$ instance.} \label{sec:cc-res-lam00}
\end{figure}

\begin{figure}[h!]
\begin{center}\vspace{-20pt} $
\begin{array}{ll}
\hspace{-30pt}
\includegraphics[width=0.35\textwidth, angle=-90]{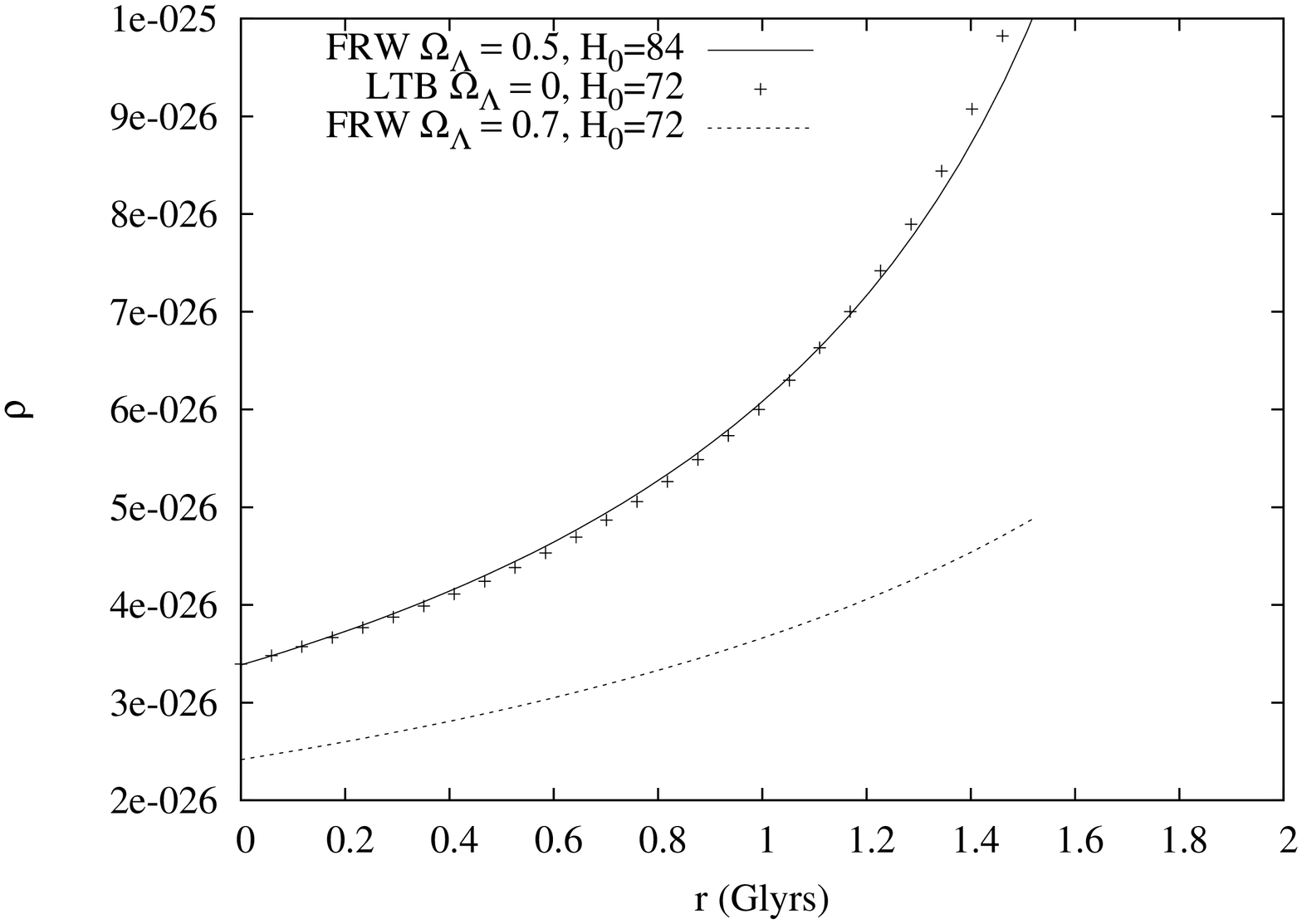} &
\hspace{-20pt}\includegraphics[width=0.35\textwidth,
angle=-90]{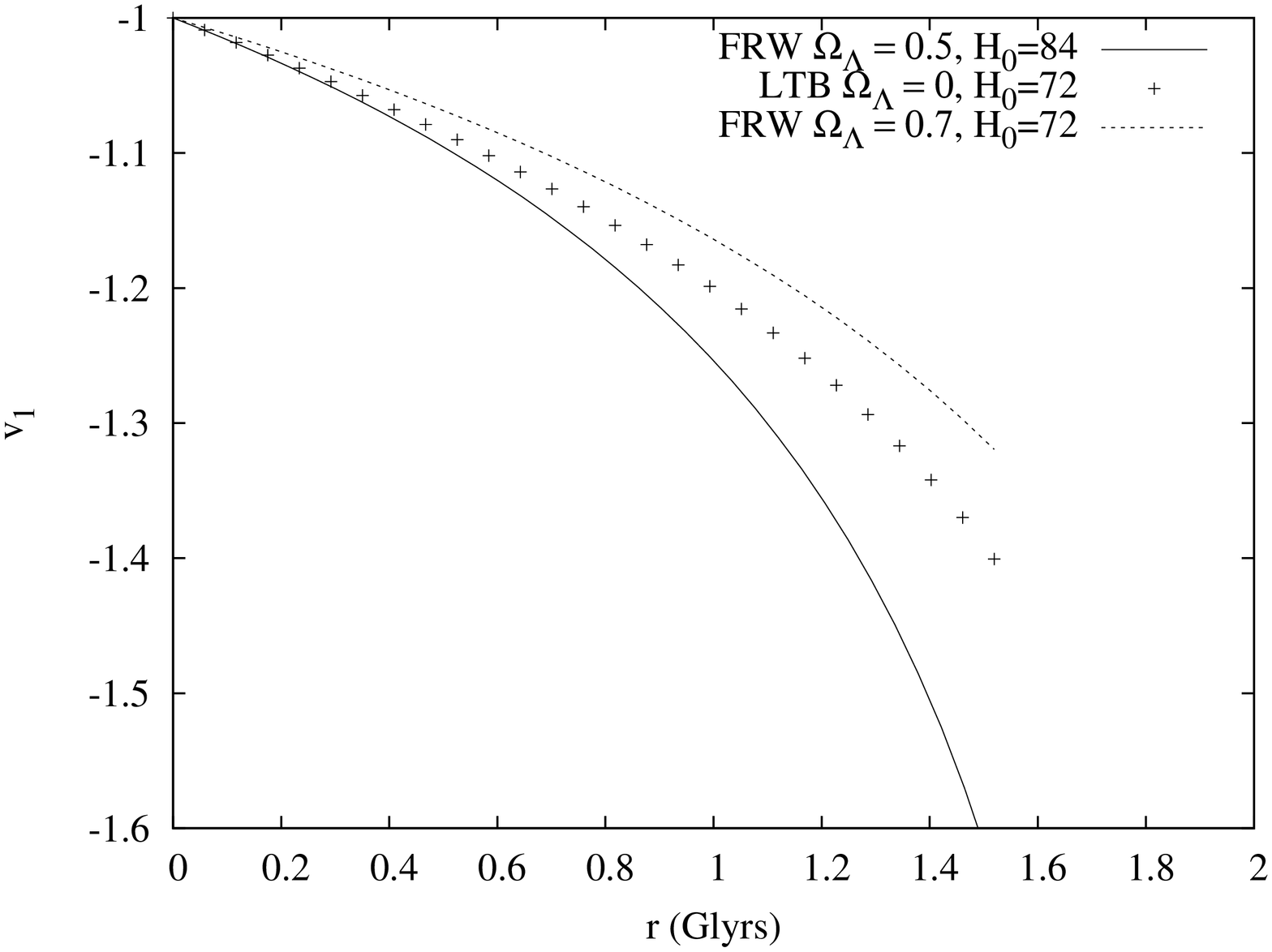}
\end{array}$
\end{center} \vspace{-20pt} \caption{Density distribution and covariant velocity
on the past null cones: LTB with $\Omega_\Lambda=0$ and FLRW with
$\Omega_\Lambda=0.7$ at $t_0-8$Gyrs and the best fit FLRW
$\Omega_\Lambda=0.5$ instance.} \label{sec:cc-res-lam05}
\end{figure}

\begin{figure}[h!]
\begin{center}$
\begin{array}{ll}
\hspace{-30pt}
\includegraphics[width=0.35\textwidth, angle=-90]{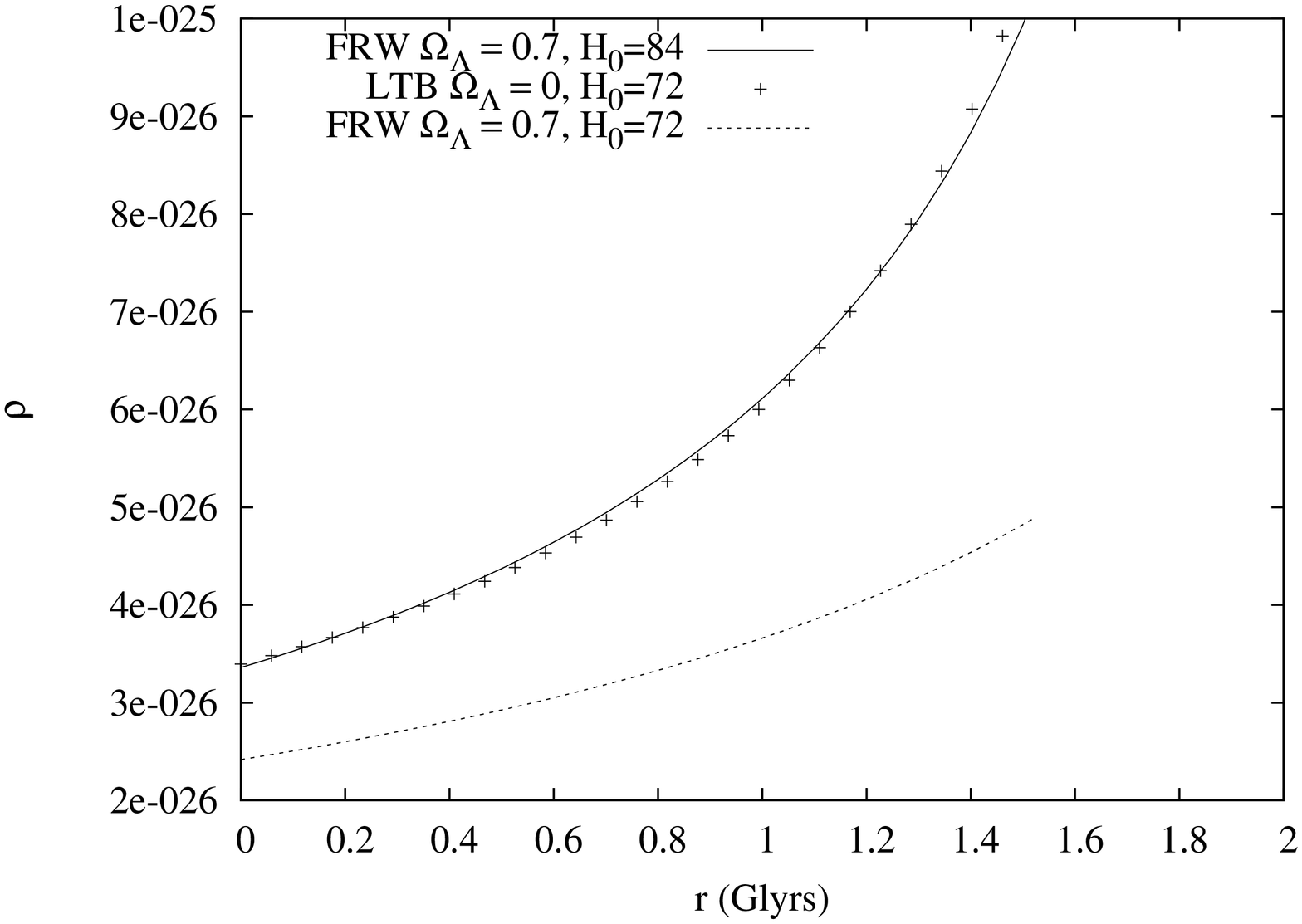} &
\hspace{-20pt}\includegraphics[width=0.35\textwidth,
angle=-90]{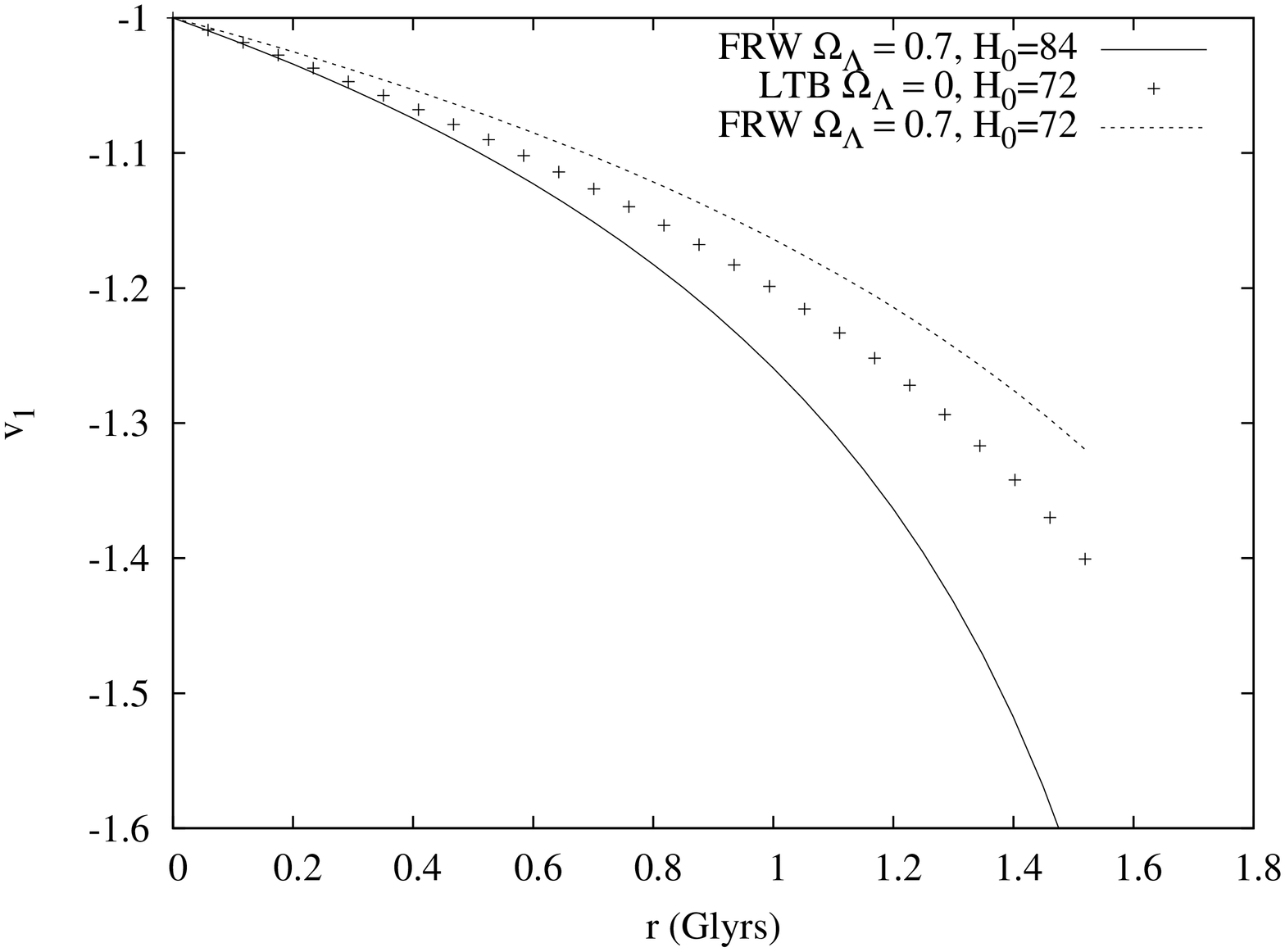}
\end{array}$
\end{center} \vspace{-20pt} \caption{Density distribution and covariant velocity
on the past null cones: LTB with $\Omega_\Lambda=0$ and FLRW with
$\Omega_\Lambda=0.7$ at $t_0-8$Gyrs and the best fit FLRW
$\Omega_\Lambda=0.7$ instance.} \label{sec:cc-res-lam07}
\end{figure}

\begin{figure}[h!]
\begin{center}$
\begin{array}{ll}
\hspace{-30pt}
\includegraphics[width=0.35\textwidth, angle=-90]{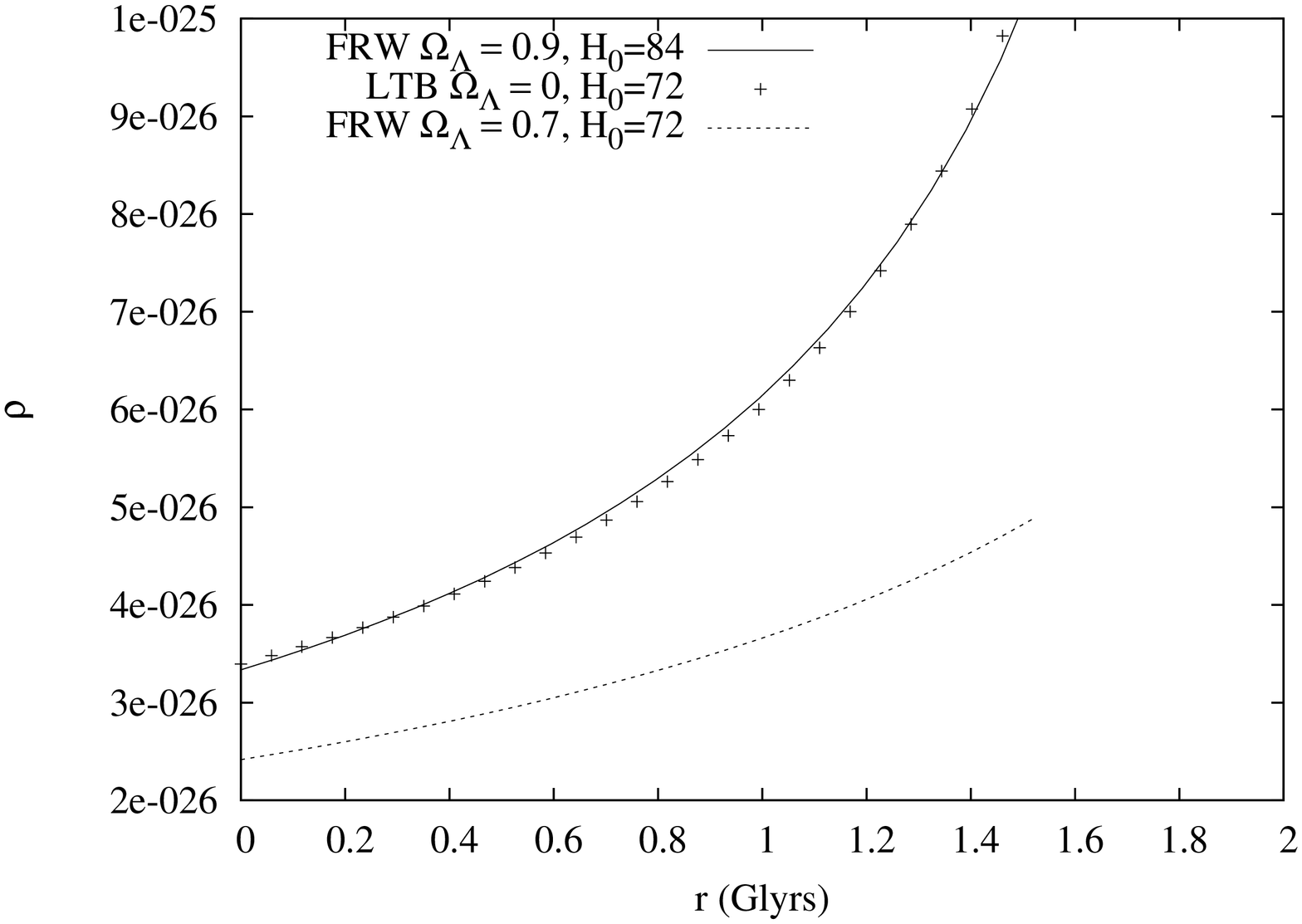} &
\hspace{-20pt}\includegraphics[width=0.35\textwidth,
angle=-90]{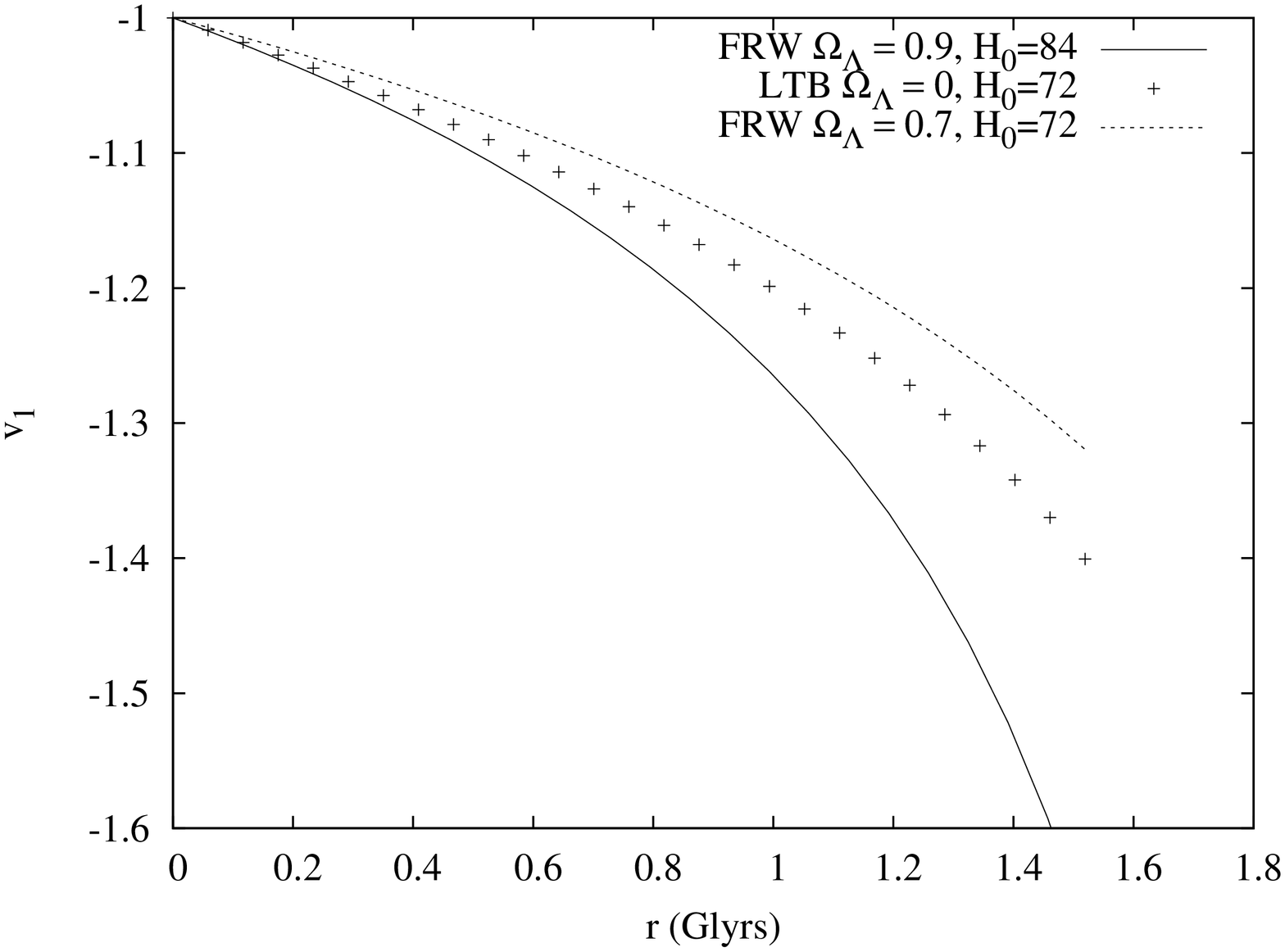}
\end{array}$
\end{center} \vspace{-20pt} \caption{Density distribution and covariant velocity
on the past null cones: LTB with $\Omega_\Lambda=0$ and FLRW with
$\Omega_\Lambda=0.7$ at $t_0-8$Gyrs and the best fit FLRW
$\Omega_\Lambda=0.9$ instance.} \label{sec:cc-res-lam09}
\end{figure}

The models investigated are summarised in table \ref{sec:cc-tab-frw}
and detailed results are shown in the figures \ref{sec:cc-res-lam00}
to \ref{sec:cc-res-lam09}. From table \ref{sec:cc-tab-frw} it can be
seen that the matching instances are from universes with different
ages and matching takes place at different times in their
evolutions. From the detailed results, it becomes apparent that
matching both the radial matter distribution and expansion through
$\rho$ and $v_1$ is an unlikely proposition. Applying the matching
procedure to $\rho$ causes $v_1$ to move away from the $\Lambda$CDM
data while $\rho$ moves away when $v_1$ is matched. This then
suggests that the past null cones do not represent a $\Lambda$CDM
model. This is an illustration of similar conclusions found by Yoo,
Kai \& Nakao \cite{yoo08}.

An interesting correlation is that the values of $H_0$ and
$t_{match}$ are very close for all the models and it might be
possible that they are in fact the same. A formal correlation was
however not done since the qualitative behaviour of the covariant
velocities rules out a reasonable match in any of the models. The
plots are presented in figures (\ref{sec:cc-res-lam00}) to
(\ref{sec:cc-res-lam09}) which reveal a surprising similarity in
the $v_1$ plots.

~\\
\begin{table}[h!]
\begin{center}
\begin{tabular}{ l c c c c}
\hline
$\Omega_{\Lambda}$ & $H_0$ & Age ($t_0$) & Density matching time  & Figure\\
 & (Mpc s$^{-1}$ km$^{-1}$) & (Gyrs) & ($t_{match}$) (Gyrs) & \\
\hline
$0$ & $86$ & $8.34$ & $4.44$ & \ref{sec:cc-res-lam00}\\
$0.5$ & $84$ & $10.66$ & $4.48$ & \ref{sec:cc-res-lam05}\\
$0.7$ & $84$ & $12.36$ & $4.45$ & \ref{sec:cc-res-lam07}\\
$0.9$ & $84$ & $16.39$ & $4.43$ & \ref{sec:cc-res-lam09}\\
\hline
\end{tabular}
\end{center}
\caption{$\Lambda$CDM models used to match the LTB density profile.}
\label{sec:cc-tab-frw}
\end{table}

\section{Conclusion}
\label{s-Conc}
In this paper we demonstrated how to use the characteristic
formalism of numerical relativity to investigate problems in
observational cosmology. For this purpose, a CIVP code was developed
and it was shown that the code is second order accurate and stable
for selected LTB models in the region before the PNC starts to
refocus. By doing a numerical experiment with LTB vs. $\Lambda$CDM
data, it was demonstrated that although the initial values of the
two models can correspond on the current past null cone, the
histories of the two models are distinctly different. The density of
the LTB model rises significantly more quickly indicating a much younger
universe, possibly too young. The result that past null cones evolved
from an initial LTB null cone cannot be matched with a flat
$\Lambda$CDM model has an important implication: While in our
current epoch the LTB vs. $\Lambda$CDM ambiguity is difficult to
disentangle, this is a feature of the Universe's current state and
not its past. \emph{In other words, if the Universe is inhomogeneous
without a cosmological constant, not only is the observer in a
privileged position (near a central point), he also lives in a
specific time where the Universe can appear to be either LTB or
$\Lambda$CDM}. It would be interesting to investigate extending this
result to non-flat models.

\acknowledgments
This work was supported by the National Research Foundation, South Africa.

\bibliographystyle{unsrt}
\bibliography{paper}

\end{document}